\documentclass[aps,prl,reprint,superscriptaddress]{revtex4-2}
\usepackage{amsmath,amssymb,graphicx,booktabs}
\usepackage{xurl}
\ifdefined\DeclareUnicodeCharacter
  \DeclareUnicodeCharacter{2212}{\ensuremath{-}}
  \DeclareUnicodeCharacter{00B0}{\ensuremath{^\circ}}
\fi
\providecommand{\mathscr}{\mathcal}
\DeclareFontFamily{OT1}{cmss}{}
\DeclareFontShape{OT1}{cmss}{m}{n}{<->cmss10}{}
\usepackage{color}
\definecolor{publicationlink}{rgb}{0.12,0.29,0.42}
\usepackage[unicode,colorlinks=true,allcolors=publicationlink]{hyperref}

\graphicspath{{./}}
\hypersetup{pdfauthor={Weican Yang, Xiangdong Xu, Cheng Li, Keyu Lu},pdftitle={Geometry-controlled zeros of resonant Kelvin-wave conversion in four spatial dimensions}}
\begin{document}
\title{Geometry-controlled zeros of resonant Kelvin-wave conversion in four spatial dimensions}
\author{Weican Yang}
\email[Corresponding author: ]{weicanyang@outlook.com}
\affiliation{Yukawa Institute for Theoretical Physics, Kyoto University, Kyoto 606-8502, Japan}
\author{Xiangdong Xu}
\noaffiliation
\author{Cheng Li}
\noaffiliation
\author{Keyu Lu}
\noaffiliation
\begin{abstract}
We show that geometry can suppress resonant four-wave conversion of Kelvin waves on a vortex surface in four spatial dimensions. Perpendicular incoming waves eliminate the entire local-area contribution, exposing a finite interaction mediated by the surrounding fluid. Reflection allows the incoming wave-number ratio to vary at exact resonance, revealing a sign-changing zero of this fluid coupling at \(\lambda_*\simeq1.469\). Uniform matching of the Gross--Pitaevskii (GP) four-wave coupling to the fluid result establishes microscopic zeros that approach this ratio at long wavelengths. Numerical calculations also show the corresponding sign change. Independent calculations in separate resonant geometries connect the coupling sign to the phase of coherently generated waves and the frequency ordering of reflection-even and reflection-odd modes. The surrounding flow thus determines whether an exactly resonant conversion channel is present at four-wave order.
\end{abstract}
\maketitle
\raggedbottom
Kelvin waves connect vortex motion to small-scale energy transfer and sound emission in superfluids {[}1--9{]}. Their spectrum determines the kinematics of resonant conversion, while the nonlinear interaction sets its strength and phase. Direct excitation and observation in helium {[}10--11{]}, GP simulations of reconnection and sound emission {[}12{]}, and classical-vortex experiments {[}13{]} have brought these dynamics into reach. Local tension supplies the dominant long-wavelength contribution, while the surrounding circulation field produces finite interactions that are essential in three-dimensional Kelvin-wave theory {[}6--9,14--15{]}. Removing the local coupling would expose these interactions directly and reveal which resonant conversions the surrounding flow can sustain.

A vortex surface in four spatial dimensions {[}16--23{]} provides this separation. Its normal core remains two-dimensional, as for an ordinary vortex line, while two tangent directions allow perpendicular incoming wavevectors and nontrivial four-wave conversion. On a strictly convex filament branch, two-to-two four-wave resonance only permutes momenta {[}1--2,6--9{]}. On the surface, perpendicular incidence eliminates the complete local-area coupling {[}24{]}. Reflecting the incoming pair about its total momentum preserves the paired magnitudes and hence exact resonance for the actual isotropic Kelvin dispersion. The wave-number ratio can therefore vary the interaction without introducing frequency mismatch.

Here we show that the surrounding flow restores conversion for generic reflected quartets, yet produces a zero where the finite coupling changes sign. This zero of the nonlocal interaction survives in the complete GP four-wave coefficient, with its ratio approaching \(\lambda_*\simeq1.469\) at long wavelengths. Finite-domain calculations follow the sign reversal, while separate coherent-mixing and probe calculations show how the coupling sign enters generated-wave phases and reflection-resolved frequencies. The contrast between local-area and full-flow predictions exposes nonlinear vortex dynamics that the Kelvin spectrum alone does not determine.

\begin{figure*}[t]
\centering
\includegraphics[width=0.98\textwidth]{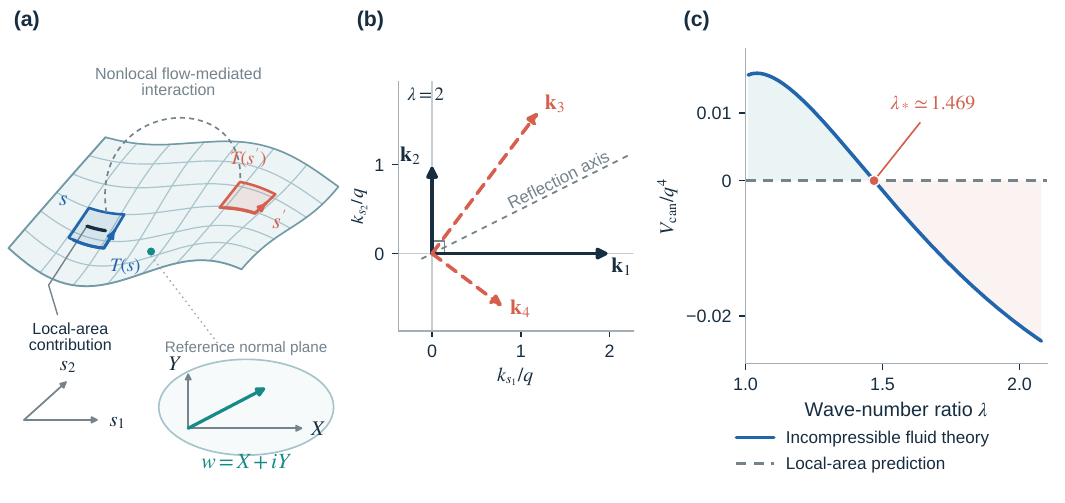}
\caption{Geometry exposes a conversion zero at exact resonance. (a) Schematic surface projection with local-area energy and a flow-mediated pair interaction (dashed link, not a trajectory); independent normal displacements \(w=X+iY\). (b) Quartet (3) at \(\lambda=2\), in units of \(q\); solid/dashed arrows denote incoming/outgoing waves. Reflection about the total-momentum axis preserves paired magnitudes. (c) The local-area prediction and incompressible fluid coefficient \(V_{\rm can}/q^4=f(\lambda)\) in the same canonical normalization. The finite flow coupling changes sign at \(\lambda_*\simeq1.469\); its microscopic continuation is established below.}
\end{figure*}

\begin{figure*}[t]
\centering
\includegraphics[width=0.98\textwidth]{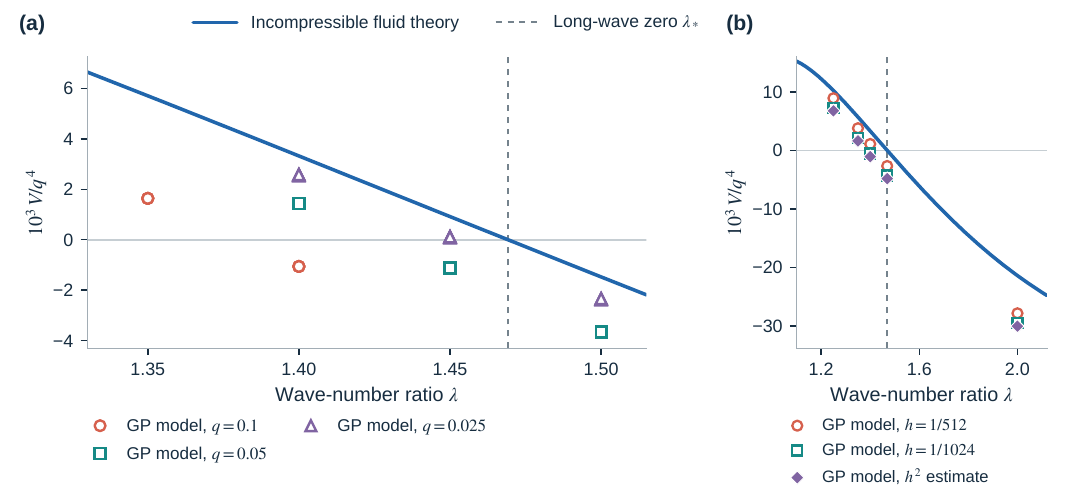}
\caption{GP coupling for quartets (3), plotted as \(10^3V/q^4\). (a) Radial collocation at three wave numbers, \(V=g_{\rm GP}^{(R)}\); overlapping symbols include radius and tolerance variations. The dashed guide marks the long-wave fluid zero \(\lambda_*\). (b) Finite-volume comparison at \(q=0.1,R=256\), with \(V=g_{\rm GP}^{(R,h)}\); diamonds assume leading \(h^2\) error. The smooth curve in both panels is incompressible fluid theory, \(V=q^4f(\lambda)\). Here \(R\) is the normal cutoff radius and \(h\) the radial spacing; numerical controls are given in Sec. S5.}
\end{figure*}

\begin{figure*}[t]
\centering
\includegraphics[width=0.98\textwidth]{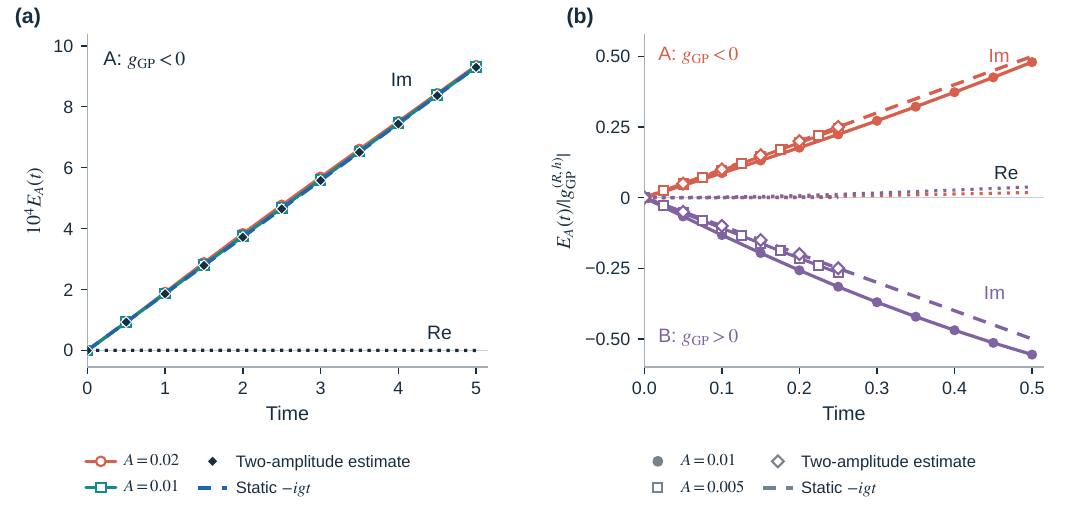}
\caption{Coherent GP mixing. The phase-corrected receiver increment divided by \(A^3\) is \(E_A\) {[}Eq. (9){]}. (a) Shape A at \(q=0.4\), \(A=0.02,0.01\), plotted as \(10^4E_A\). (b) A/B at \(q=0.2\), normalized by \(|g_{\rm GP}^{(R,h)}|\); circles/squares denote \(A=0.01,0.005\). The same-grid coefficient is \(g_{\rm GP}^{(R,h)}<0\) for A, giving positive imaginary growth, and positive for B, giving negative growth. Solid and dotted curves show the imaginary and real parts, respectively; diamonds denote two-amplitude estimates, and dashed curves show the independent same-grid predictions \(-ig_{\rm GP}^{(R,h)}t\). A/B are separate resonant geometries from the reflected scan in Fig. 2 {[}25, Sec. S3{]}.}
\end{figure*}

Represent the deformed vortex surface by \(F(s)=(X,Y,s_1,s_2)\), with tangent coordinates \(s=(s_1,s_2)\) and real normal displacements \(X,Y\) {[}Fig. 1(a){]}. Write \(w=X+iY\), and let \(J=X_{s_1}Y_{s_2}-X_{s_2}Y_{s_1}\) be the normal-displacement Jacobian. The gradient and subscripts denote tangent derivatives. The true area density is \(\sqrt{1+|\nabla w|^2+J^2}\). For tension \(\tau\), its expansion gives the quadratic wave energy and leading four-wave interaction {[}17--18,25--27{]},

\[
H_2=\frac{\tau}{2}\int|\nabla w|^2d^2s,\qquad
H_4=-\frac{\tau}{8}\int|\nabla w\cdot\nabla w|^2d^2s.
\tag{1}
\]

Here \(H_2\) and \(H_4\) are the quadratic and quartic parts of the local-area Hamiltonian, and \(\tau\) is energy per actual area. In \(w(s)=\sum_kw_ke^{ik\cdot s}\), \(k\) is a tangent wavevector and \(w_k\) its Fourier amplitude. For four distinct modes, \(V_{w,12;34}\) denotes the coefficient of \(w_1w_2w_3^*w_4^*\) per projected area, with \(w_i=w_{k_i}\) and its conjugate term written separately. Extracting this term from \(H_4\), with incoming modes 1,2 and outgoing modes 3,4, yields

\[
V^{\mathrm{area}}_{w,12;34}
=-\frac{\tau}{2}(k_1\cdot k_2)(k_3\cdot k_4),
\qquad k_1+k_2=k_3+k_4.
\tag{2}
\]

Perpendicular incidence therefore removes the complete area contribution for any tension, including its finite renormalization. The Gram determinant \(J^2\) is essential to this cancellation {[}24--25{]}. To vary the remaining interaction while preserving resonance, reflect the incoming pair about its total momentum.

Take \(q=|k_2|>0\) as the reference incoming wave number and \(\lambda=|k_1|/|k_2|>0\), \(\lambda\ne1\), as the ratio. Choose

\[
\begin{aligned}
k_1&=q(\lambda,0),\qquad k_2=q(0,1),\\
k_3&=\frac{q}{1+\lambda^2}\bigl(\lambda(\lambda^2-1),2\lambda^2\bigr),\\
k_4&=\frac{q}{1+\lambda^2}\bigl(2\lambda,1-\lambda^2\bigr).
\end{aligned}
\tag{3}
\]

The construction gives \(k_1+k_2=k_3+k_4\), \(|k_1|=|k_3|\) and \(|k_2|=|k_4|\) {[}Fig. 1(b){]}. Thus it preserves frequency matching for any isotropic Kelvin dispersion \(\omega(|k|)\). Changing \(\lambda\) changes the individual frequencies while preserving their resonant relation. Angle resonances {[}28{]} and zeros of resonant wave interactions {[}29--30{]} provide precedents. Related geometric structure appears directly in the water-wave action {[}31{]}. Here the entire area contribution has already vanished, isolating the coupling carried by the surrounding circulation.

That flow is described by the four-dimensional Green function \((4\pi^2|x|^2)^{-1}\), where \(x\) is the ambient separation. The tangent bivector \(T=F_{s_1}\wedge F_{s_2}\) represents the oriented surface-area element. For bulk mass density \(\rho\), circulation \(\Gamma\) and ambient-distance core cutoff \(a_c>0\), its energy is {[}16--18,25{]}

\[
\begin{aligned}
H_{a_c}&=K_E\iint\frac{T(s)\cdot T(s')}{|F(s)-F(s')|^2+a_c^2}\,d^2s\,d^2s',\\
&\qquad K_E=\frac{\rho\Gamma^2}{8\pi^2}.
\end{aligned}
\tag{4}
\]

Both the relative orientation and separation of surface elements contribute to the four-wave coupling. Expanding them together cancels the power divergences and separates an area-like logarithm from a finite, momentum-dependent interaction. For fixed wavevectors with nonzero relevant transfers, the limit \(a_c|k_i|\to0\) gives

\[
\begin{aligned}
\frac{V_{w,12;34}}{\pi K_E}
&=B[\ln(a_c/2)+\gamma_E]+\mathcal F(k_i)+o(1),\\
B&=(k_1\cdot k_2)(k_3\cdot k_4).
\end{aligned}
\tag{5}
\]

Here \(\gamma_E\) is Euler's constant and \(\mathcal F\) is the finite remainder, containing momentum-transfer logarithms and nonlogarithmic terms {[}25, Sec. S1{]}. All logarithms use one inverse-length reference; the finite coefficient at \(B=0\) is independent of it. The logarithm carries exactly the geometric factor of the local-area vertex. Perpendicular incidence removes it, together with the leading local tension change induced by a short-distance radial regulator. The remaining finite flow interaction is therefore leading: for healing length \(\xi\) and \(\Lambda=\ln[1/(q\xi)]\gg1\), the generic \(O(q^4\Lambda)\) coupling becomes \(O(q^4)\) in the selected geometry.

Write \(V_w\) for the displacement vertex \(V_{w,12;34}\). For canonical fluid amplitudes \(a_k=\sqrt{\rho\Gamma/2}\,w_k\), the vertex is \(V_{\rm can}=4V_w/(\rho\Gamma)^2\). Set \(\rho=1\), \(\Gamma=2\pi\). On (3), the leading finite vertex is \(V_{\rm can}=q^4 f(\lambda)\), where, with \(\Sigma_\lambda=1+\lambda^2\) and \(d_\lambda=\lambda^2-1\),

\[
\begin{aligned}
8\pi f(\lambda)&=d_\lambda^2\ln|d_\lambda|-\lambda^4\ln(\lambda^2)\\
&\quad+2\lambda^2\ln\Sigma_\lambda+\frac{\lambda^2(\lambda^4-6\lambda^2+1)}{\Sigma_\lambda^2}.
\end{aligned}
\tag{6}
\]

No nonlinear coefficient is fitted. In particular, \(f(7/5)>0>f(3/2)\), and the first fluid zero for \(\lambda>1\) is \(\lambda_*\simeq1.469\) {[}Fig. 1(c){]}. This sign change concerns the complete finite flow term, rather than a cancellation between an adjustable tension and a nonlocal correction.

Local-area theory predicts zero conversion for every quartet in (3). The full flow restores conversion at generic ratios and produces a zero of its own. Whether a microscopic vortex core preserves that zero is the next physical question.

We consider the GP model {[}32--33{]}, setting the bulk density, chemical potential, particle mass and \(\hbar\) to one:

\[
i\partial_t\psi=[-\nabla^2/2+|\psi|^2-1]\psi.
\tag{7}
\]

Here \(\psi\) is the condensate field, \(t\) is time and \(\nabla\) is the four-dimensional gradient; \(\xi=1/\sqrt2\). We follow the Kelvin branch that approaches transverse core translation at long wavelengths. For the exact regular positive radial unit vortex on the infinite plane normal to the surface {[}34{]}, this is the lowest branch with positive Bogoliubov (Krein) norm in the \(m=-1\) angular sector measured relative to the background winding. The particle and hole radial components \(u,v\) are normalized by \(\int(|u|^2-|v|^2)d^2x_\perp=1\), with \(x_\perp\) the normal coordinates {[}35--36{]}.

Let \(a_i^{\rm GP}\) be mode amplitudes with this normalization. Matching the core displacement relates these GP amplitudes to the canonical fluid amplitudes at long wavelengths. Let \(g_{\rm GP}(q,\lambda)\) be the full four-wave coupling coefficient for the process \(1+2\leftrightarrow3+4\) in their Hamiltonian. It contains the direct term and all three intermediate responses, \((1,2),(1,-3),(2,-3)\), with negative labels denoting particle--hole conjugates {[}25{]}. These responses are second-order vortex-field deformations generated by pairs of the participating modes. With the mode phases fixed by the displacement of the same vortex core, reflection exchanges incoming and outgoing modes and makes \(g_{\rm GP}\) real.

The core could add a four-wave contribution of the same order as the finite flow interaction. At long wavelengths, its leading local energy follows the full induced area, whose selected four-wave coefficient vanishes for (3). We match the Kelvin modes and all three pair-generated deformations to the circulation field by comparing complete actions, including the symplectic term and moving core boundary. Stationarity makes response errors enter the action quadratically, leaving a uniform \(o(q^4)\) difference over an interval containing the fluid zero {[}25, Sec. S4{]}.

The full GP coupling therefore retains the conversion zero. For sufficiently small \(q>0\),

\[
\begin{gathered}
\lim_{q\to0}\sup_{\lambda\in I}
\left|q^{-4}g_{\rm GP}(q,\lambda)-f(\lambda)\right|=0,
\qquad I=[7/5,3/2],\\
g_{\rm GP}(q,\lambda_q)=0,\qquad \lambda_q\in I,\\
\lambda_q\longrightarrow \lambda_*\simeq1.469\quad(q\to0).
\end{gathered}
\tag{8}
\]

Uniform matching preserves the strict endpoint signs, while the decrease of \(f\) fixes the limiting ratio of the microscopic zeros. Reflection makes the coefficient real in the common displacement convention. The cancellation is channel selective: diagonal frequency shifts and higher-order processes remain allowed.

Figure 2 follows the change of sign into the GP model. As \(q\) decreases from \(0.1\) to \(0.025\), the collocation scans move the sampled reversal from \([1.35,1.40]\) to \([1.45,1.50]\), towards \(\lambda_*\). Independent finite-volume calculations give the same first bracket. The finite-radius data display the long-wave trend; the near-zero sensitivity and numerical controls are given in {[}25, Sec. S5{]}. In the fluid limit, the short-time generated action has a quadratic minimum as the ratio passes through the zero, accompanied by a phase reversal {[}25, Sec. S3{]}.

With fixed pump and seed phases, changing the coupling sign reverses the phase of the leading generated wave. To read out this sign we use two other resonant geometries, A and B, with perpendicular incoming wavevectors of equal magnitude and opposite GP couplings {[}25, Sec. S3{]}. Unlike the reflected quartets in Eq. (3), A/B choose the direction of \(k_3-k_4\) and set its magnitude by resonance on the computed GP branch. Modes 1 and 2 are pumps, mode 3 is a seed and mode 4 is the receiver. Each mode has unit Krein norm, with its phase fixed by the same core displacement. The pump and seed scheme parallels matter-wave mixing {[}37--39{]}. The initial input amplitudes are equal and real, \(a_{1,2,3}^{\rm GP}(0)=A\).

For linear frequencies \(\omega_i=\omega(|k_i|)\), the leading receiver amplitude is

\[
\begin{aligned}
a_4^{\rm GP}(t)e^{i\omega_4t}-a_4^{\rm GP}(0)
&=-i g_{\rm GP}^{(R,h)}a_1^{\rm GP}(0)\\
&\quad\times a_2^{\rm GP}(0)a_3^{\rm GP}(0)^*t+\cdots .
\end{aligned}
\tag{9}
\]

The independently calculated \(g_{\rm GP}^{(R,h)}\) uses the same grid as the trajectory. The initial field includes the inputs and their quadratic responses. Dividing the generated increment in (9) by \(A^3\) defines \(E_A\). The weak-input trajectories follow the predicted cubic response and reverse the generated-wave phase between A and B (Fig. 3) {[}25, Sec. S3{]}.

% Reflection-resolved probe measurement.
The coupling sign also sets a frequency ordering fixed by physical reflection symmetry. In the fluid normalization, let \(N_i=|a_{k_i}|^2\) be action per tangent area and \(g_{\rm fl}=V_{\rm can}\) the coupling. For this measurement, modes 2 and 4 are equal-phase pumps related by reflection, with \(N_2=N_4=N\); modes 1 and 3 are probes with initial actions \(N_1=P,N_3=0\). If the probe pair is isolated and depletion is negligible, reflection makes its diagonal shifts equal, while the symmetric and antisymmetric combinations \((a_1\pm a_3)/\sqrt2\) acquire shifts \(\pm g_{\rm fl}N\) about a common frequency. 

\noindent\begin{minipage}{\columnwidth}
Their even and odd frequencies, \(\Omega_{\rm even}\) and \(\Omega_{\rm odd}\), thus obey

\[
\begin{aligned}
\Omega_{\rm even}-\Omega_{\rm odd}&=2g_{\rm fl}N,\\
N_3(t)&=P\sin^2(|g_{\rm fl}|Nt),\qquad N_1(0)=P.
\end{aligned}
\tag{10}
\]
\end{minipage}\par

Here \(P\) is the initial probe action density, small compared with the pump action \(N\). Reflection exchanges the probes; the frequency order of its even and odd modes reverses with \(g_{\rm fl}\). \mbox{Weak-pump GP spectra} recover the coupling through this signed splitting {[}25, Sec. S3{]}. The probe-isolation and pump-depletion conditions are discussed there, including the role of sound coupling {[}3,14,40{]}.

The same finite flow contribution is also required for consistency between Kelvin dispersion and scattering in a separate soft-momentum limit {[}25, Sec. S6{]}.

Local-area theory leaves all the reflected quartets uncoupled; the surrounding flow restores conversion and selects zeros of its own. The first fluid zero survives the microscopic core in the long-wavelength GP four-wave coefficient. Its coupling sign enters coherent generation and reflection-resolved probe frequencies, giving nonlinear vortex spectroscopy a direct target {[}6,14--15,24,37--39{]}. The surrounding medium therefore determines both the strength of the selected leading interaction and whether its resonant conversion is present at all.

The immediate question is how this channel-selective structure competes with other resonant processes and nonlinear broadening to shape angular redistribution {[}28,41{]}. For physical three-dimensional vortex lines, the corresponding object is the complete resonant six-wave amplitude, including virtual four-wave processes {[}6--9{]}. Searching it for related cancellations would connect this mechanism to Kelvin-wave transport along ordinary vortices. More generally, geometric cancellations offer a calculable route to separating local defect tension from interactions mediated by bulk fields {[}14--15,24{]}, complementing the symmetry constraints used in effective theories of strings and branes {[}42--43{]}.

\hypertarget{references}{%
\section{References}\label{references}}

\begingroup
\raggedright
\setlength{\leftmargini}{1.8em}
\begin{enumerate}
\setlength{\itemsep}{0pt}\setlength{\parsep}{0pt}
\linespread{0.97}\selectfont
\def\labelenumi{[\arabic{enumi}]}
\item
  W. F. Vinen, M. Tsubota, and A. Mitani, \emph{Kelvin-wave cascade on a vortex in superfluid \({}^{4}\mathrm{He}\) at a very low temperature}, \href{https://doi.org/10.1103/PhysRevLett.91.135301}{Phys. Rev.~Lett. \textbf{91}, 135301 (2003)}.
\item
  E. Kozik and B. Svistunov, \emph{Kelvin-wave cascade and decay of superfluid turbulence}, \href{https://doi.org/10.1103/PhysRevLett.92.035301}{Phys. Rev.~Lett. \textbf{92}, 035301 (2004)}; \href{https://arxiv.org/abs/cond-mat/0308193}{arXiv:cond-mat/0308193}.
\item
  E. Kozik and B. Svistunov, \emph{Vortex-phonon interaction}, \href{https://doi.org/10.1103/PhysRevB.72.172505}{Phys. Rev.~B \textbf{72}, 172505 (2005)}; \href{https://arxiv.org/abs/cond-mat/0505020}{arXiv:cond-mat/0505020}.
\item
  B. V. Svistunov, \emph{Superfluid turbulence in the low-temperature limit}, \href{https://doi.org/10.1103/PhysRevB.52.3647}{Phys. Rev.~B \textbf{52}, 3647 (1995)}.
\item
  C. Nore, M. Abid, and M. E. Brachet, \emph{Kolmogorov Turbulence in Low-Temperature Superflows}, \href{https://doi.org/10.1103/PhysRevLett.78.3896}{Phys. Rev.~Lett. \textbf{78}, 3896 (1997)}.
\item
  J. Laurie, V. S. L'vov, S. Nazarenko, and O. Rudenko, \emph{Interaction of Kelvin waves and nonlocality of energy transfer in superfluids}, \href{https://doi.org/10.1103/PhysRevB.81.104526}{Phys. Rev.~B \textbf{81}, 104526 (2010)}.
\item
  V. S. L'vov and S. Nazarenko, \emph{Spectrum of Kelvin-wave turbulence in superfluids}, \href{https://doi.org/10.1134/S002136401008014X}{JETP Lett. \textbf{91}, 428--434 (2010)}; \href{https://arxiv.org/abs/0911.2065}{arXiv:0911.2065}.
\item
  L. Boué, R. Dasgupta, J. Laurie, V. L'vov, S. Nazarenko, and I. Procaccia, \emph{Exact solution for the energy spectrum of Kelvin-wave turbulence in superfluids}, \href{https://doi.org/10.1103/PhysRevB.84.064516}{Phys. Rev.~B \textbf{84}, 064516 (2011)}.
\item
  A. W. Baggaley and J. Laurie, \emph{Kelvin-wave cascade in the vortex filament model}, \href{https://doi.org/10.1103/PhysRevB.89.014504}{Phys. Rev.~B \textbf{89}, 014504 (2014)}; \href{https://arxiv.org/abs/1204.4034}{arXiv:1204.4034}.
\item
  E. Fonda, D. P. Meichle, N. T. Ouellette, S. Hormoz, and D. P. Lathrop, \emph{Direct observation of Kelvin waves excited by quantized vortex reconnection}, \href{https://doi.org/10.1073/pnas.1312536110}{Proc. Natl. Acad. Sci. U.S.A. \textbf{111}, 4707--4710 (2014)}.
\item
  Y. Minowa, Y. Yasui, T. Nakagawa, S. Inui, M. Tsubota, and M. Ashida, \emph{Direct excitation of Kelvin waves on quantized vortices}, \href{https://doi.org/10.1038/s41567-024-02720-9}{Nat. Phys. \textbf{21}, 233--238 (2025)}.
\item
  M. Leadbeater, T. Winiecki, D. C. Samuels, C. F. Barenghi, and C. S. Adams, \emph{Sound Emission due to Superfluid Vortex Reconnections}, \href{https://doi.org/10.1103/PhysRevLett.86.1410}{Phys. Rev.~Lett. \textbf{86}, 1410 (2001)}; \href{https://arxiv.org/abs/cond-mat/0009060}{arXiv:cond-mat/0009060}.
\item
  J. Barckicke, C. Gissinger, and E. Falcon, \emph{Experimental Evidence of Kelvin-Wave Turbulence along a Vortex Core}, \href{https://doi.org/10.1103/t3bt-m431}{Phys. Rev.~Lett. \textbf{137}, 134001 (2026)}.
\item
  B. Horn, A. Nicolis, and R. Penco, \emph{Effective string theory for vortex lines in fluids and superfluids}, \href{https://doi.org/10.1007/JHEP10(2015)153}{J. High Energy Phys. \textbf{10} (2015), 153}; \href{https://arxiv.org/abs/1507.05635}{arXiv:1507.05635}.
\item
  S. S. Gubser, R. Nayar, and S. Parikh, \emph{Strings, vortex rings, and modes of instability}, \href{https://doi.org/10.1016/j.nuclphysb.2015.01.005}{Nucl. Phys. B \textbf{892}, 156--180 (2015)}; \href{https://arxiv.org/abs/1408.2246}{arXiv:1408.2246}.
\item
  P. M. Goldbart and F. Bora, \emph{Quantized vortices and superflow in arbitrary dimensions: Structure, energetics and dynamics}, \href{https://doi.org/10.1088/1751-8113/42/18/185001}{J. Phys. A \textbf{42}, 185001 (2009)}.
\item
  B. N. Shashikanth, \emph{Vortex dynamics in \(\mathbb{R}^4\)}, \href{https://doi.org/10.1063/1.3673800}{J. Math. Phys. \textbf{53}, 013103 (2012)}.
\item
  B. Khesin, \emph{Symplectic structures and dynamics on vortex membranes}, \href{https://doi.org/10.17323/1609-4514-2012-12-2-413-434}{Mosc. Math. J. \textbf{12}, 413--434 (2012)}; \href{https://arxiv.org/abs/1201.5914}{arXiv:1201.5914}.
\item
  B. McCanna and H. M. Price, \emph{Curved vortex surfaces in four-dimensional superfluids. I. Unequal-frequency double rotations}, \href{https://doi.org/10.1103/PhysRevA.110.013325}{Phys. Rev.~A \textbf{110}, 013325 (2024)}.
\item
  H. A. J. Middleton-Spencer, B. McCanna, D. Proment, and H. M. Price, \emph{Interactions and reconnections of four-dimensional quantum vortices}, \href{https://doi.org/10.1103/nmz7-7ym2}{Phys. Rev.~Fluids \textbf{11}, 084701 (2026)}.
\item
  W.-c.~Yang, \emph{Non-equilibrium Dynamics and Universality of 4D Quantum Vortices and Turbulence}, \href{https://arxiv.org/abs/2504.04409v3}{arXiv:2504.04409v3} (2025).
\item
  B. McCanna and H. M. Price, \emph{Superfluid vortices in four spatial dimensions}, \href{https://doi.org/10.1103/PhysRevResearch.3.023105}{Phys. Rev.~Research \textbf{3}, 023105 (2021)}.
\item
  B. McCanna and H. M. Price, \emph{Curved vortex surfaces in four-dimensional superfluids. II. Equal-frequency double rotations}, \href{https://doi.org/10.1103/PhysRevA.110.013326}{Phys. Rev.~A \textbf{110}, 013326 (2024)}.
\item
  M. A. Mojahed and T. Brauner, \emph{Nonrelativistic effective field theories with enhanced symmetries and soft behavior}, \href{https://doi.org/10.1007/JHEP03(2022)086}{J. High Energy Phys. \textbf{03} (2022), 086}; \href{https://arxiv.org/abs/2201.01393}{arXiv:2201.01393}.
\item
  \hyperlink{supplemental-material}{Supplemental Material}: geometry and the finite nonlocal interaction; GP modes and microscopic matching; static coefficients, coherent mixing and even/odd probe spectra; angular conversion and the soft-momentum relation; technical proofs and numerical methods.
\item
  J. Huang and D. Tataru, \emph{Local well-posedness of the skew mean curvature flow for small data in \(d\geq2\) dimensions}, \href{https://doi.org/10.1007/s00205-023-01952-y}{Arch. Ration. Mech. Anal. \textbf{248}, 10 (2024)}; \href{https://arxiv.org/abs/2202.10632}{arXiv:2202.10632}.
\item
  Z. Li, \emph{Global and local theory of skew mean curvature flows}, \href{https://doi.org/10.1007/s12220-021-00735-9}{J. Geom. Anal. \textbf{32}, 34 (2022)}; \href{https://arxiv.org/abs/2010.14011}{arXiv:2010.14011}.
\item
  E. Kartashova, S. Nazarenko, and O. Rudenko, \emph{Resonant interactions of nonlinear water waves in a finite basin}, \href{https://doi.org/10.1103/PhysRevE.78.016304}{Phys. Rev.~E \textbf{78}, 016304 (2008)}.
\item
  N. S. Ussembayev, \emph{Non-interacting gravity waves on the surface of a deep fluid}, \href{https://doi.org/10.48550/arXiv.1903.10854}{arXiv:1903.10854} (2019).
\item
  N. Arkani-Hamed, F. Calisto, N. Ussembayev, W. W. Zhao, and Z. Zhou, \emph{Surface Water Wave Scattering and the Hydrotope}, \href{https://doi.org/10.48550/arXiv.2606.28280}{arXiv:2606.28280} (2026).
\item
  Q. Cao, S. He, J. Jing, and Q. Li, \emph{The Hydrotope in the Water-Wave Action}, \href{https://doi.org/10.48550/arXiv.2608.26881}{arXiv:2608.26881} (2026).
\item
  E. P. Gross, \emph{Structure of a quantized vortex in boson systems}, \href{https://doi.org/10.1007/BF02731494}{Nuovo Cimento \textbf{20}, 454--477 (1961)}.
\item
  L. P. Pitaevskii, \emph{Vortex lines in an imperfect Bose gas}, \href{https://jetp.ras.ru/cgi-bin/dn/e_013_02_0451.pdf}{Sov. Phys. JETP \textbf{13}, 451--454 (1961)}.
\item
  R.-M. Hervé and M. Hervé, \emph{Étude qualitative des solutions réelles d'une équation différentielle liée à l'équation de Ginzburg-Landau}, \href{https://doi.org/10.1016/S0294-1449(16)30182-2}{Ann. Inst. H. Poincaré Anal. Non Linéaire \textbf{11}, 427--440 (1994)}.
\item
  D. A. Takahashi, M. Kobayashi, and M. Nitta, \emph{Nambu-Goldstone modes propagating along topological defects: Kelvin and ripple modes from small to large systems}, \href{https://doi.org/10.1103/PhysRevB.91.184501}{Phys. Rev.~B \textbf{91}, 184501 (2015)}; \href{https://arxiv.org/abs/1501.01874}{arXiv:1501.01874}.
\item
  E. Papoutsis, N. Apfel, and N. Navon, \emph{Core-bound waves on a Gross-Pitaevskii vortex}, \href{https://arxiv.org/abs/2603.05505}{arXiv:2603.05505} (2026).
\item
  L. Deng, E. W. Hagley, J. Wen, M. Trippenbach, Y. B. Band, P. S. Julienne, J. E. Simsarian, K. Helmerson, S. L. Rolston, and W. D. Phillips, \emph{Four-wave mixing with matter waves}, \href{https://doi.org/10.1038/18395}{Nature \textbf{398}, 218--220 (1999)}.
\item
  M. Trippenbach, Y. B. Band, and P. S. Julienne, \emph{Theory of four-wave mixing of matter waves from a Bose-Einstein condensate}, \href{https://doi.org/10.1103/PhysRevA.62.023608}{Phys. Rev.~A \textbf{62}, 023608 (2000)}; \href{https://arxiv.org/abs/cond-mat/0002119}{arXiv:cond-mat/0002119}.
\item
  Q. Yang, J. T. Seo, S. Creekmore, D. A. Temple, P. Ye, C. Bonner, M. Namkung, S. S. Jung, and J. H. Kim, \emph{Nonlinear phase mismatch and optimal input combination in atomic four-wave mixing in Bose-Einstein condensates}, \href{https://doi.org/10.1103/PhysRevA.67.013603}{Phys. Rev.~A \textbf{67}, 013603 (2003)}.
\item
  S. Endlich and A. Nicolis, \emph{The incompressible fluid revisited: vortex-sound interactions}, \href{https://doi.org/10.48550/arXiv.1303.3289}{arXiv:1303.3289} (2013).
\item
  S. Chibbaro, G. Dematteis, and L. Rondoni, \emph{4-wave dynamics in kinetic wave turbulence}, \href{https://doi.org/10.1016/j.physd.2017.09.001}{Physica D \textbf{362}, 24--59 (2018)}; \href{https://arxiv.org/abs/1611.08030}{arXiv:1611.08030}.
\item
  P. Cooper, S. Dubovsky, V. Gorbenko, A. Mohsen, and S. Storace, \emph{Looking for Integrability on the Worldsheet of Confining Strings}, \href{https://doi.org/10.1007/JHEP04(2015)127}{J. High Energy Phys. \textbf{04} (2015), 127}; \href{https://arxiv.org/abs/1411.0703}{arXiv:1411.0703}.
\item
  O. Aharony and Z. Komargodski, \emph{The Effective Theory of Long Strings}, \href{https://doi.org/10.1007/JHEP05(2013)118}{J. High Energy Phys. \textbf{05} (2013), 118}; \href{https://arxiv.org/abs/1302.6257}{arXiv:1302.6257}.
% REVTeX stores frontmatter author notes separately from this manual reference list.
% Emit them with its native formatter here, then prevent a second auto bibliography.
\makeatletter
\let\@TBN@opr\present@bibnote
\@FMN@list
\global\let\@FMN@list\@empty
\global\let\auto@bib\@empty
\makeatother
\end{enumerate}
\endgroup

% ======================================================================
% Supplemental Material (single-column layout)
% ======================================================================
\clearpage
\onecolumngrid
\begingroup
\setcounter{secnumdepth}{0}
\setcounter{section}{0}
\setcounter{subsection}{0}
\setcounter{equation}{0}
\setcounter{figure}{0}
\setcounter{table}{0}
\renewcommand{\theequation}{S\arabic{equation}}
\renewcommand{\thefigure}{S\arabic{figure}}
\renewcommand{\thetable}{S\arabic{table}}
% Keep hyperlink destinations distinct from those in the main text.
\renewcommand{\theHsection}{supplement.\arabic{section}}
\renewcommand{\theHequation}{supplement.\arabic{equation}}
\renewcommand{\theHfigure}{supplement.\arabic{figure}}
\renewcommand{\theHtable}{supplement.\arabic{table}}
\hypertarget{supplemental-material}{}
\pdfbookmark[0]{Supplemental Material}{supplemental-material-bookmark}
\begin{center}
{\large\bfseries Supplemental Material:\par
Geometry-controlled zeros of resonant Kelvin-wave conversion in four spatial dimensions\par}
\vspace{1em}
Weican Yang\textsuperscript{1,*}, Xiangdong Xu, Cheng Li, and Keyu Lu\par
\vspace{4pt}
{\small\textsuperscript{1} Yukawa Institute for Theoretical Physics, Kyoto University, Kyoto 606-8502, Japan\par}
\vspace{4pt}
{\small\textsuperscript{*} Corresponding author: \href{mailto:weicanyang@outlook.com}{weicanyang@outlook.com}\par}
\end{center}
\medskip

\raggedbottom
\hypertarget{s1.-finite-nonlocal-interaction}{%
\section{S1. Finite nonlocal interaction}\label{s1.-finite-nonlocal-interaction}}

Write the vortex surface as \(F=(X,Y,s_1,s_2)\), with complex normal displacement \(w=X+iY\). Its graph symplectic form {[}S1--S3{]} is exactly \(\varpi=\rho\Gamma\int\delta X\wedge\delta Y\,d^2s\): the tangent identity block in the ambient volume determinant removes all slope corrections. Thus \(a_k=\sqrt{\rho\Gamma/2}\,w_k\) obeys \(i\dot a_k=\partial H/\partial a_k^*\). Vertices below are per projected tangent area: they are the coefficients of the distinct-mode monomial \(w_1w_2w_3^*w_4^*\), with its conjugate counted separately.

With \(S=|\nabla w|^2\), \(J=X_{s_1}Y_{s_2}-X_{s_2}Y_{s_1}\), and \(G=I+\nabla X\nabla X^T+\nabla Y\nabla Y^T\), the area identity gives
\[
\det G=1+S+J^2,\quad
H_4^{\rm area}=-\frac\tau8\int|\nabla w\cdot\nabla w|^2d^2s,\quad
V_w^{\rm area}=-\frac\tau2(k_1\cdot k_2)(k_3\cdot k_4).
\tag{S1}\label{eq:area}
\]

Here \(|\nabla w\cdot\nabla w|^2=S^2-4J^2\); the two orderings of each pair supply the factor four. Perpendicular incoming waves remove the full area coefficient, including the finite part of the tension.

For \(T=F_{s_1}\wedge F_{s_2}\), the four-dimensional Green function gives
\[
\begin{aligned}
H_{a_c}&=K_E\iint\frac{T(s)\cdot T(s')}{|F(s)-F(s')|^2+a_c^2}\,d^2s\,d^2s',\qquad K_E=\frac{\rho\Gamma^2}{8\pi^2},\\
H_{4,a_c}&=K_E\iint\left(\frac{JJ'}D-\frac{A_\nabla|\Delta w|^2}{D^2}+\frac{|\Delta w|^4}{D^3}\right)d^2s\,d^2s'.
\end{aligned}
\tag{S2}\label{eq:fluid-energy}
\]

Here \(D=|s-s'|^2+a_c^2\), \(\Delta w=w(s)-w(s')\), and \(A_\nabla=\operatorname{Re}[\nabla w(s)\cdot\nabla w(s')^*]\). All three terms use one ambient radial regulator. For a localized integrable change \(\Delta g\), freezing the slope and setting \(t=G^{1/2}(s-s')\) yields
\[
\frac{\delta H_{\rm loc}}{d^2s}=K_E\det G\int\Delta g(R^TGR)d^2R
=K_E\sqrt{\det G}\int\Delta g(t^2)d^2t=\delta\tau\sqrt{\det G}.
\tag{S3}\label{eq:regulator-area}
\]

Curvature introduces additional derivatives, so the leading perpendicular \(q^4\) coefficient is independent of this hydrodynamic cutoff.

To specify all multiplicities, set \(u=x_1e^{ik_1\cdot s}+x_2e^{ik_2\cdot s}\), \(\bar u=y_3e^{-ik_3\cdot s}+y_4e^{-ik_4\cdot s}\), and let \([M]\) extract \(x_1x_2y_3y_4\). Define \(\mathcal J=\tfrac i2(\partial_1u\partial_2\bar u-\partial_2u\partial_1\bar u)\), \(\mathcal A(s,s')=\tfrac12[\nabla u(s)\cdot\nabla\bar u(s')+\nabla\bar u(s)\cdot\nabla u(s')]\), and \(\mathcal D=(u-u')(\bar u-\bar u')\). Write
\[
[M]\{\mathcal J(0)\mathcal J(-R),\mathcal A(0,-R)\mathcal D(0,-R),\mathcal D(0,-R)^2\}
=\sum_Q\{j_Q,b_Q,d_Q\}e^{iQ\cdot R}.
\tag{S4}\label{eq:vertex-generator}
\]

For \(p=|Q|\), \(L_p=\ln(a_cp/2)+\gamma_E\), the transforms of \(D^{-n}\) are
\[
\begin{aligned}
F_1&=-2\pi L_p+o(1),&
F_2&=\frac\pi{a_c^2}+\frac{\pi p^2}2(L_p-\tfrac12)+o(1),\\
F_3&=\frac\pi{2a_c^4}-\frac{\pi p^2}{8a_c^2}-\frac{\pi p^4}{32}(L_p-\tfrac34)+o(1).
\end{aligned}
\tag{S5}\label{eq:transforms}
\]

These follow from \(F_1=2\pi K_0(a_cp)\) and \(F_{n+1}=-\partial_{a_c^2}F_n/n\). The vertex is \(K_E\sum_Q(j_QF_1-b_QF_2+d_QF_3)\). Evaluation at \(R=0\) and \(\mathcal D(0,-R)^2=O(R^4)\) give \(\sum b_Q=\sum d_Q=\sum p^2d_Q=0\). Momentum conservation then gives
\[
\begin{gathered}
B=-2\sum j_Q-\tfrac12\sum b_Qp^2-\tfrac1{32}\sum d_Qp^4
=(k_1\cdot k_2)(k_3\cdot k_4),\\
\frac{V_w}{\pi K_E}=B[\ln(a_c/2)+\gamma_E]+\mathcal F+o(1),\\
\mathcal F=\sum_Q\left[-2j_Q\ln p-\tfrac12b_Qp^2\ln p-\tfrac1{32}d_Qp^4\ln p+\tfrac14b_Qp^2+\tfrac3{128}d_Qp^4\right].
\end{gathered}
\tag{S6}\label{eq:finite-vertex}
\]

Changing the reference scale in the logarithms shifts only the term proportional to \(B\). At zero transfer, \(p^n\ln p\) has its continuous value for \(n>0\); channels with \(j_0\ne0\), requiring another infrared prescription, are excluded. Canonical normalization gives \(V_{\rm can}=4V_w/(\rho\Gamma)^2\). In GP units \(\rho=1,\Gamma=2\pi\), this is \(V_w/\pi^2\), and \(a=\sqrt\pi w\).

Take \(k_1=q(\lambda,0)\), \(k_2=q(0,1)\), and reflect both about their sum:
\[
k_3=\frac q{1+\lambda^2}(\lambda(\lambda^2-1),2\lambda^2),\qquad
k_4=\frac q{1+\lambda^2}(2\lambda,1-\lambda^2).
\tag{S7}\label{eq:reflected}
\]

This preserves the sum and paired magnitudes, giving resonance for every isotropic dispersion; the modes are distinct for \(\lambda\ne1\). Substitution in (S4)--(S6) yields \(V_{\rm can}=q^4f(\lambda)\), where
\[
\begin{aligned}
f(\lambda)&=\frac{F_\perp(\lambda^2,1)}{2\pi},\\
F_\perp(a,b)&=\frac{(a-b)^2\ln|a-b|-a^2\ln a-b^2\ln b+2ab\ln(a+b)}4
+\frac{ab(a^2-6ab+b^2)}{4(a+b)^2}.
\end{aligned}
\tag{S8}\label{eq:fluid-function}
\]

The logarithm associated with the squared transfer \(4ab/(a+b)\) cancels between the three contributions; their polynomial parts give the rational term. On \(I=[7/5,3/2]\), \(2\pi f(1.4)=0.02086676533\), \(2\pi f(1.5)=-0.00926111594\), and the fluid zero is \(\lambda_*=1.469117113\ldots\). Section S4 proves its microscopic persistence. The ordering \(q^4\) versus generic \(q^4\ln[1/(q\xi)]\) assumes comparable transfers and subsonic Kelvin motion.

\hypertarget{s2.-definition-of-the-microscopic-gp-coefficient}{%
\section{S2. Definition of the microscopic GP coefficient}\label{s2.-definition-of-the-microscopic-gp-coefficient}}

Use \(i\partial_t\psi=-\Delta\psi/2+(|\psi|^2-1)\psi\), with bulk density, mass and \(\hbar\) equal to one and \(\xi=1/\sqrt2\). The exact positive vortex is \(\psi_0=f_0(r)e^{i\vartheta}\), where
\[
-\tfrac12(f_0''+r^{-1}f_0'-r^{-2}f_0)+(f_0^2-1)f_0=0,\qquad f_0(0)=0,\quad f_0(\infty)=1.
\tag{S9}\label{eq:vortex}
\]

For \(\psi=e^{i\vartheta}(f_0+\delta)\), the nonlinearities are \(\mathcal N_2=f_0(\delta^2+2\delta\delta^*)\), \(\mathcal N_3=\delta^2\delta^*\) {[}S4--S6{]}. For relative angular label \(m\) and physical angular order \(\ell\), define
\[
\begin{gathered}
D_\ell(k)=-\tfrac12(\partial_r^2+r^{-1}\partial_r-\ell^2/r^2)+|k|^2/2+2f_0^2-1,\quad
L_m(k)=\begin{pmatrix}D_{1+m}&f_0^2\\-f_0^2&-D_{1-m}\end{pmatrix},\\
L_{-1}(k_i)Z_i=\omega_iZ_i,\quad Z_i=(u_i,v_i)^T,\quad
\int_0^\infty(|u_i|^2-|v_i|^2)\,2\pi r\,dr=1,\quad\eta=\operatorname{diag}(1,-1).
\end{gathered}
\tag{S10}\label{eq:kelvin-definition}
\]

For radial vectors, use the ordinary Hermitian inner product \(\langle X,Y\rangle=\int_0^\infty X^\dagger Y\,d\mu\), \(d\mu=2\pi r\,dr\), and \(\|X\|^2=\langle X,X\rangle\); a subscript restricts the radial domain. The Krein pairing is instead \(\langle X,\eta Y\rangle\), as in (S10). We select the variationally lowest positive-Krein Kelvin branch with continuous real phase and positive translation overlap. Both Nambu components enter the field:
\[
\delta_1=\sum_i[a_i u_i e^{i\chi_i}+a_i^*v_i^*e^{-i\chi_i}],\quad
\chi_i=k_i\cdot s-\vartheta-\omega_it,\quad
\delta=\epsilon_{\rm amp}\delta_1+\epsilon_{\rm amp}^2\delta_2+\cdots.
\tag{S11}\label{eq:physical-first-field}
\]

\par\addvspace{\medskipamount}
\noindent\begin{minipage}{\textwidth}
\textbf{Table S1.} Pair-generated responses contributing to the receiver \(a_1a_2a_3^*\). The field \(\delta_2\) contains these responses and their conjugates.
\par\smallskip\centering
\begin{tabular}{@{}lrll@{}}
\toprule
Pair & \(m\) & Momentum \(K\) & Frequency \(\nu\) \\
\midrule
\((1,2)\) & \(-2\) & \(k_1+k_2\) & \(\omega_1+\omega_2\) \\
\((1,-3)\) & \(0\) & \(k_1-k_3\) & \(\omega_1-\omega_3\) \\
\((2,-3)\) & \(0\) & \(k_2-k_3\) & \(\omega_2-\omega_3\) \\
\bottomrule
\end{tabular}
\end{minipage}\par\medskip

A negative leg reverses momentum, frequency and angular label and applies \(\mathsf C(p,q)=(q^*,p^*)\). For distinct signed legs \(e_i=(p_i,q_i)^T\),
\[
\mathbf T_{ij}=2f_0\binom{p_ip_j+p_iq_j+p_jq_i}{q_iq_j+p_iq_j+p_jq_i},\quad
P_{ij}Z^{(2)}_{ij}=-\mathbf T_{ij},\quad
P_{ij}=\eta L_{m_i+m_j}(k_i+k_j)-(\omega_i+\omega_j)\eta.
\tag{S12}\label{eq:pair-response}
\]

Both leg orderings are included; the evolution source is \(\eta\mathbf T_{ij}\), and \(\mathbf T_{-i,-j}=\mathsf C\mathbf T_{ij}\). The full cubic source and signed projection define
\[
\begin{aligned}
S_{3,u}&=[a_1a_2a_3^*]\{\delta_1^2\delta_1^*+2f_0(\delta_1\delta_2+\delta_1\delta_2^*+\delta_2\delta_1^*)\},\\
g_{\rm GP}&=\int_0^\infty(u_4^*S_{3,u}-v_4^*S_{3,v})\,2\pi r\,dr.
\end{aligned}
\tag{S13}\label{eq:gp-coefficient}
\]

The lower source is minus the conjugate upper evolution source at the opposite monomial. There is no additional factorial. Replacing infinity by \(R\), then discretizing with radial spacing \(h\), defines \(g_{\rm GP}^{(R)}\), \(g_{\rm GP}^{(R,h)}\). Both retain the three responses. A pole would require additional dynamical variables or a specified scattering prescription; S4 excludes poles uniformly for small \(q\) in the reflected family. Reflection exchanges \(1\leftrightarrow3\), \(2\leftrightarrow4\), so the action and unique responses make \(g_{\rm GP}\) real in this phase convention.

\hypertarget{s3.-coherent-generation-and-reflection-parity-readout}{%
\section{S3. Coherent generation and reflection-parity readout}\label{s3.-coherent-generation-and-reflection-parity-readout}}

Figure 3 uses \(k_1=q(1,0)\), \(k_2=q(0,1)\), \(k_{3,4}=(k_1+k_2)/2\pm k_{\rm out}\hat n\), with \(\hat n=(1,0)\) for A and \((\cos(-\pi/8),\sin(-\pi/8))\) for B. Solving for \(k_{\rm out}\) enforces computed-branch resonance, typically to \(O(10^{-12})\). These differ from Fig. 2's reflected family. Initialize three equal real unit-Krein inputs \(A\), their conjugates and quadratic responses; then
\[
E_A=\frac{a_4(t)e^{i\omega_4t}-a_4(0)}{A^3},\qquad
E_{\rm weak}=\frac{4E_{A/2}-E_A}{3}\simeq-ig_{\rm GP}^{(R,h)}t.
\tag{S14}\label{eq:coherent-readout}
\]

The subtraction removes the dressed initial projection; extrapolation assumes a leading \(A^2\) correction. The independent static coefficient uses the trajectory's grid (Sec. S5); its sign fixes the generated phase relative to the inputs {[}S7{]}.

\par\addvspace{\medskipamount}
\noindent\begin{minipage}{\textwidth}
\textbf{Table S2.} Same-grid mixing comparisons. The discrepancy is \(|E_{\rm weak}+igt|/(|g|t)\); \(g=g_{\rm GP}^{(R,h)}\).
\par\smallskip\centering
\begin{tabular}{@{}lrrlrr@{}}
\toprule
Shape, \(q\) & \(h\) & \(g\) & Inputs \(A\) & \(t\) & Discrepancy \\
\midrule
A, 0.4 & 1/64 & \(-1.86026\times10^{-4}\) & 0.02/0.01 & 5 & 0.0496056\% \\
A, 0.2 & 1/256 & \(-1.01321\times10^{-5}\) & 0.01/0.005 & 0.25 & 0.183891\% \\
B, 0.2 & 1/256 & \(+4.29074\times10^{-6}\) & 0.01/0.005 & 0.25 & 0.391365\% \\
\bottomrule
\end{tabular}
\end{minipage}\par\medskip

The same-grid comparisons test the weak-input, short-time readout. At \(q=0.4,t=5\), doubling \(A\) gives receiver-gain ratio \(8.03306+0.00394i\). Halving the timestep changes the main \(A=0.01\) result by 0.0231880\% at \(t=1\), and B at \(q=0.2,t=0.25\) by 0.332477\%. The main half-timestep control ends before \(t=5\), and these checks do not determine phase-grid, continuum or long-time errors.

For the reflected fluid family, \(g_{\rm fl}=q^4f(\lambda)\), \(f(\lambda_*)=0\), \(f'_*<0\). Equal canonical fluid inputs \(A_{\rm fl}\) give the leading weak-input, short-time increment \(b_{4,\rm fl}=-ig_{\rm fl}A_{\rm fl}^3t\), with action
\[
|b_{4,\rm fl}|^2\sim A_{\rm fl}^6t^2q^8(f'_*)^2(\lambda-\lambda_*)^2.
\tag{S15}\label{eq:fluid-quadratic-minimum}
\]

This quadratic minimum accompanies a \(\pi\) phase reversal. It is a fluid prediction; A/B trajectories do not scan that zero.

For equal-phase reflected pumps of action \(N\), the isolated probes \(\mathbf a=(a_1,a_3)^T\) obey \(i\dot{\mathbf a}=(\Omega_h I+Ng_{\rm fl}\sigma_x)\mathbf a\). Reflection diagonalizes this matrix into \((a_1\pm a_3)/\sqrt2\), giving main Eq. (10). The common self- and cross-phase shift {[}S7--S8{]} remains at a conversion zero. Rephasing transforms the Hamiltonian and reflection together, preserving parity ordering. An isolated smooth GP doublet gives \(\Delta\sigma=\sigma_{\rm even}-\sigma_{\rm odd}=2Ng_{\rm GP}^{(R,h)}+O(N^2)\), taking \(N\to0\) at fixed \(q\), without cell-area or factorial factors. The calculated pumps are \(p_\pm=\mathbf u\pm\mathbf v\), probes \(2\mathbf u\pm\mathbf v\), with \(\mathbf u\perp\mathbf v\), \(|p_\pm|=q\). For \(d=p_-\cdot p_+/q^2\), the high/low squared ratio is \((5+3d)/2\): \(d=-1/3\) is perpendicular, \(\lambda=\sqrt2\); \(d=-0.2\) is acute.

\par\addvspace{\medskipamount}
\noindent\begin{minipage}{\textwidth}
\textbf{Table S3.} Independent \(g_{\rm GP}^{(R,h)}\) and measured \(\Delta\sigma/(2N)\) at \(q=0.2\), in \(10^{-6}\). Columns use seed \(A\); the quotient uses achieved action, with target \(N=A^2\).
\par\smallskip\centering
\begin{tabular}{@{}rrrrr@{}}
\toprule
\(d\) & Independent \(g\) & \(A=0.025\) & \(A=0.050\) & \(A=0.100\) \\
\midrule
\(-1/3\) & \(-6.79877\) & \(-6.80372\) & \(-6.81851\) & \(-6.87683\) \\
\(-0.2\) & \(+33.69845\) & \(+33.69251\) & \(+33.67466\) & \(+33.60416\) \\
\bottomrule
\end{tabular}
\end{minipage}\par\medskip

The calculated spectra display signed ordering, and the quotient approaches the independent coefficient as \(N\) decreases. At \(h=1/256\), static coefficients retain both signs and shift to \(-9.63273\times10^{-6}\), \(+3.08579\times10^{-5}\). The perpendicular fluid coefficient is positive at long wavelengths, unlike this finite-\(q\) GP value.

An exchange interpretation requires probe action \(P\ll N\), an isolated doublet, and depletion-induced detuning and sound linewidths \(\ll N|g|\). Reflection neither guarantees stability nor equates imaginary self-energies. These finite-grid comparisons do not establish continuum values or certify frequency errors and doublet isolation; outgoing sound loss, long-time exchange and near-zero spectra have not been measured. Resolution becomes stricter as \(g\to0\).

\hypertarget{s4.-uniform-microscopic-matching-and-the-conversion-zero}{%
\section{S4. Uniform microscopic matching and the conversion zero}\label{s4.-uniform-microscopic-matching-and-the-conversion-zero}}

\textbf{Theorem.} For the infinite normal plane, the S2 branch and S1 reflected quartets,
\[
\sup_{\lambda\in I}|q^{-4}g_{\rm GP}(q,\lambda)-f(\lambda)|
\le3e_\rho(q)^2+e_A(q)+e_F(q)\longrightarrow0.
\tag{S16}\label{eq:master-bound}
\]

For sufficiently small \(q>0\), a real zero exists in \(I\), and every such zero converges to \(\lambda_*\); fixed-\(q\) uniqueness is not asserted.

Lemmas 1 and 2 control the response operators and establish the Kelvin-mode normalization and tail bounds. Lemma 3 constructs a common trial field and bounds its response residuals. In Lemma 4, stationarity makes these errors quadratic, and comparison of the complete actions bounds the GP--fluid difference. Lemma 5 then identifies the finite fluid vertex.

Set \(I=[7/5,3/2]\), \(L=1+|\ln q|\), \(R_q=(qL^6)^{-1}\). Then \(R_q\to\infty\) and \(qR_q=L^{-6}\to0\): the matching region lies outside the fixed GP core and below the wavelength. The power six is a technical choice; for example, it makes \(q^{-4}q^6L^5R_q^2=L^{-7}\) in Lemma 4. Constants are uniform on \(I\); finite logarithmic exponents \(P\) may change only beside positive powers of \(q\). At fixed \(q>0\), take the outer radius \(R_{\rm IR}\to\infty\), then remove the auxiliary fluid tube radius \(\epsilon\) or kernel cutoff \(a_c\) in the common subtractions, before taking \(q\to0\). The GP healing length stays fixed; \(R_q\) is a separate matching radius. Let \(\operatorname{Pol}_4\) extract \(a_1a_2a_3^*a_4^*\), with coefficients \(\Gamma_{S,G},\Gamma_{S,T},\Gamma_{S,F}\) in the exact GP, trial GP and trial fluid actions per tangent area and time. The response, action and fluid-identification errors are
\[
\begin{gathered}
e_\rho=q^{-2}\sup_I\max_{ij}\|\rho_{ij}\|_{P_{ij}^{-1}},\quad
e_A=q^{-4}\sup_I|\Gamma_{S,T}-\Gamma_{S,F}|,\\
e_F=\sup_I|-q^{-4}\Gamma_{S,F}-f|,\qquad
\|\rho\|_{P^{-1}}^2=\langle\rho,P^{-1}\rho\rangle.
\end{gathered}
\tag{S17}\label{eq:error-definitions}
\]

Here \(\rho_{ij}\) is the trial second-order response residual defined after (S31). Its inverse-response norm measures the error relevant to the stationary action: \(\|\rho_{ij}\|_{P_{ij}^{-1}}=o(q^2)\) gives an \(o(q^4)\) action error in Lemma 4.

\hypertarget{lemma-1.-positive-response-operators}{%
\subsection{Lemma 1. Positive response operators}\label{lemma-1.-positive-response-operators}}

Put \(\mathsf T_\ell=-(\partial_r^2+r^{-1}\partial_r)/2+\ell^2/(2r^2)\), \(E_K=K^2/2\), and \(c_0=\sqrt3-5/4>0\). In density/phase coordinates \(x=(u+v)/\sqrt2\), \(y=(u-v)/\sqrt2\), on the original coupled regular domain,
\[
\begin{gathered}
\mathsf A_0=\mathsf T_1+f_0^2-1\ge0,\qquad
\mathsf B_0=\mathsf T_1+3f_0^2-1\ge c_0,\\
P_{ij}\ge\operatorname{diag}((\mathsf B_0+E_K)/4,\mathsf A_0+E_K/3)
\quad\mbox{if }\nu^2\le c_0E_K/2 .
\end{gathered}
\tag{S18}\label{eq:response-coercivity}
\]

\emph{Proof.} The positive-vortex ground-state transform gives \(\mathsf A_0\ge0\). For \(f_{\rm lb}=r/\sqrt{r^2+2}\), \(\mathcal R_v[g]=g''+r^{-1}g'-r^{-2}g+2(1-g^2)g\) has residual \(3r^3/(r^2+2)^{5/2}>0\). The ratio \(p=f_{\rm lb}/f_0\) obeys \((rf_0^2p')'=rf_0\mathcal R_v[f_{\rm lb}]+2rf_0^4p(p^2-1)\). Zero origin flux and \(p(\infty)=1\) exclude \(p>1\). Minimizing \(1/(2r^2)+3f_{\rm lb}^2-1\) gives \(c_0\). The \(m=0\) pencil is \(\left(\begin{smallmatrix}\mathsf B_0+E_K&-\nu\\-\nu&\mathsf A_0+E_K\end{smallmatrix}\right)\); its cross term is bounded by \(3\langle x,(\mathsf B_0+E_K)x\rangle/4+4\nu^2\|y\|^2/(3c_0)\). The \(m=-2\) pencil adds a nonnegative centrifugal form. The sum, beat and static transfers have ratios
\[
K_j/q=\kappa_j,\qquad
\kappa_s=\sqrt{1+\lambda^2},\quad
\kappa_b=(\lambda^2-1)/\sqrt{1+\lambda^2},\quad
\kappa_0=2\lambda/\sqrt{1+\lambda^2}.
\tag{S19}\label{eq:transfers}
\]

They are bounded away from zero on \(I\); Lemma 2 gives \(|\nu|\le Cq^2L\). Thus (S18), invertibility and the subsonic threshold hold uniformly for small \(q\). Complementary static sectors retain separate density forces. \(\square\)

\hypertarget{lemma-2.-kelvin-normalization-and-circulation-tails}{%
\subsection{Lemma 2. Kelvin normalization and circulation tails}\label{lemma-2.-kelvin-normalization-and-circulation-tails}}

For auxiliary disk radius \(2\le b\le q^{-1/2}\), all bounds are uniform in \(b\). With \(a=|k|/q\in[1,3/2]\), \(t=(f_0'+f_0/r,f_0'-f_0/r)/(2\sqrt\pi)\), the modes obey
\[
\begin{gathered}
C_a=\sqrt\pi u_a(0)/f_0'(0)=1+O(\sqrt qL),\quad Z_a=C_at+\epsilon_a,\quad
\omega_a\le\tfrac{E_a}2\ln(1/E_a)+O(E_a),\quad E_a=a^2q^2/2,\\
\|\epsilon_a\|_{H^2(r<2b)}\le Cb^2\ln(2b)q^2L^{3/2},\quad
y_{B,a}=\frac{C_aaqK_1(aqr)}{\sqrt{2\pi}},\quad
\varepsilon_q=\sqrt qL^{3/2}+q^2L^{12}.
\end{gathered}
\tag{S20}\label{eq:mode-normalization}
\]

For a fixed core radius \(r_0\) and \(r_0\le r\le3R_q\),
\[
|\epsilon_y|+r|\epsilon_y'|\le Cq^2r\ln(2r)L^{3/2},\qquad
|\epsilon_x|+r|\epsilon_x'|\le Cq^2\ln(2r)L^{3/2}/r.
\tag{S21}\label{eq:inner-mode-errors}
\]

On \(r\asymp R_q\), \(|\partial_r^j(y_a-y_{B,a})|\le C\varepsilon_qr^{-1-j}\), \(j\le4\). With \(s_q=q+r^{-1}\), \(D_q=q^2L+r^{-2}\), \(\mathcal E=Cr^{-1}(1+qr)^2e^{-c_*qr}\), the derivative bounds below hold for \(r\ge32q^{-1/2}\); \(M\ge0\) is any fixed polynomial weight:
\[
\begin{gathered}
|\partial_r^jx_a|\le CD_qs_q^j\mathcal E\ (j\le3),\qquad
|\partial_r^jy_a|\le Cs_q^j\mathcal E\ (j\le4),\\
\sup_{r\ge q^{-1/2}}r(1+qr)^M|y_a-y_{B,a}|\le C_M\sqrt qL^{3/2}.
\end{gathered}
\tag{S22}\label{eq:tail-bounds}
\]

\emph{Proof.} The vortex obeys \(f_0'>0\), \(f_0=1-(4r^2)^{-1}+O(r^{-4})\), including differentiated tails {[}S9{]}. Since \((r^3(f_0/r)')'=-2r^2(1-f_0^2)f_0<0\), both \(t_+=f_0'+f_0/r\), \(t_-=f_0/r-f_0'\) are positive. In the flipped chart \(\mathbf U=(u,-v)=(t_+\alpha,t_-\beta)\),
\[
Q_0[\mathbf U]=\tfrac12\int(t_+^2|\alpha'|^2+t_-^2|\beta'|^2)d\mu
+\int f_0^2t_+t_-|\alpha-\beta|^2d\mu .
\tag{S23}\label{eq:translation-form}
\]

Here \(\mathsf H_0\) is the zero-tangent-momentum GP energy Hessian in this flipped chart, with quadratic form \(Q_0[\mathbf U]=\langle\mathbf U,\mathsf H_0\mathbf U\rangle\). Cutting translation off at \(E_a^{-1/2}\) gives the frequency bound at unit positive Krein norm, below the bulk threshold; hence the minimum of \(Q_0+E_a\|\mathbf U\|^2\) is attained. Component moduli lower the cooperative form, giving a strictly positive minimizer. The ground-state transform of \(\mathsf H_0+E_a-\omega_a\eta\) has only this null vector; its nonzero Krein norm excludes a Jordan vector. Variational comparison and the simple eigenprojection fix a continuous real branch.

The variational bound and a truncated \(r^2\)-weighted test give
\[
\|Z_a\|^2\le CL,\quad Q_0[\mathbf U_a]\le Cq^2L,\quad
\|rZ_a'\|^2+\|rx_a\|^2+E_a\|ry_a\|^2\le CL.
\tag{S24}\label{eq:mode-moments}
\]

Weight derivatives cost \(C\|Z_a\|^2\); the exterior massive density block and phase potential \(1/(2r^2)+O(r^{-4})\) absorb the coupling since \(\omega_a^2/E_a\to0\).

\emph{Core normalization.} For the local inverse modulo translation, the weights \(rt_+^2,rt_-^2,rf_0^2t_+t_-\) scale as \((r,r^5,r^5)\) at zero and \(r^{-1}\) at infinity. Radial Cauchy gives \(\|\mathbf U-c(t_+,t_-)\|_{r<b}\le CbQ_0^{1/2}\). For \(g\perp(t_+,t_-)\) supported below \(b\), solve \(\mathsf H_0v_g=g\) on this quotient. Outside the source, \(r(t_+^2\alpha'+t_-^2\beta')=0\) and \(d=\alpha-\beta\) satisfies \(-(pd')'+Wd=0\), with \(p=rt_+^2t_-^2/(t_+^2+t_-^2)\), \(W=2rf_0^2t_+t_-\), \(W/p\to4\). Finite energy selects the decaying difference. Removing the common limiting ratio, whose interior translation norm is \(O(\sqrt{\ln(2b)})\), gives \(\|v_g\|\le Cb^2\sqrt{\ln(2b)}\|g\|\). Duality against \((\omega_a\eta-E_a)\mathbf U_a=O_{L^2}(q^2L^{3/2})\), Cartesian elliptic estimates and center calibration prove (S20). All pairings here are Hermitian as defined in S2.

On \(A=\{1<r<2\}\), set \(\widetilde C_a=\langle t,Z_a\rangle_A/\langle t,t\rangle_A\); the flipped-ratio anchor is \(\widetilde C_a/(2\sqrt\pi)\). Equation (S20) at \(b=2\) gives \(|\widetilde C_a-C_a|\le Cq^2L^{3/2}\). The translation charge is \(\langle t,\eta t\rangle_{r<b}=f_0(b)^2\). The interior signed-charge error is at most \(C|\widetilde C_a|q\sqrt L(1+b)+Cq^2L(1+b^2)\), and (S24) bounds exterior absolute charge by \(CL/b\). Taking \(b=q^{-1/2}\) yields \(\widetilde C_a,C_a=1+O(\sqrt qL)\), with positive sign from the flipped mode.

\emph{Circulation tails.} Outside the core, \(\mathsf B_ax_a=(\omega_a+r^{-2})y_a\), \(\mathsf A_ay_a=(\omega_a+r^{-2})x_a\), where \(\mathsf B_a=-\Delta_r/2+r^{-2}+3f_0^2-1+E_a\), \(\mathsf A_a=-\Delta_r/2+r^{-2}+f_0^2-1+E_a\). At \(b=q^{-1/2}\), \(|x_a(b)|\le\delta_qy_a(b)\) and \(by_a(b)=C_a[1+O(\delta_q)]/\sqrt{2\pi}\), \(\delta_q=C\sqrt{qL}\). Massive comparison propagates the first inequality. Scalar barriers are \(y_a(b)K_{\nu_\pm}(\kappa_\pm r)/K_{\nu_\pm}(\kappa_\pm b)\), with \(\nu_\pm^2=1\pm2(\delta_q+C_0b^{-2})\), \(\kappa_\pm^2=a^2q^2\pm2\delta_q\omega_a\). Their order derivatives cost \(O(L)\), logarithmic scale derivatives \(O(1)\): \(zK_\nu'+\nu K_\nu=-zK_{\nu-1}\) cancels leading powers. This proves (S22)'s last bound.

For \(r\ge32b\), \((\mathsf B_a+j/(2r^2))(D_qs_q^j\mathcal E)\ge D_qs_q^j\mathcal E/2\) controls density derivatives through \(j=3\), interleaved with phase derivatives through order four. The commutator is \([\partial_r,\Delta_r]=-r^{-2}\partial_r\); traces on successively doubled disks are retained and exponentially absorbed. On \(r\asymp R_q\),
\[
(-\Delta_r/2+(2r^2)^{-1}+E_a)(y_a-y_{B,a})
=(\omega_a+r^{-2})x_a-[f_0^2-1+(2r^2)^{-1}]y_a
\]

has source derivatives \(O(r^{-5-j})\), \(j\le2\); interval estimates give \(\varepsilon_q\). Finally,
\[
(-\Delta_r/2+r^{-2}+3f_0^2-1)\epsilon_x
=r^{-2}\epsilon_y+\omega_ay_a-E_ax_a .
\]

Caccioppoli and the regular matrix Kato inequality give (S21) for \(\epsilon_y\); the massive supersolution \(Cq^2\ln(2r)L^{3/2}/r\) and its derivative give \(\epsilon_x\). Cartesian core estimates supply the derivative trace. Center calibration makes both physical errors quadratic at zero, regularizing the divided corrections below. \(\square\)

\hypertarget{lemma-3.-trial-fields-and-response-residuals}{%
\subsection{Lemma 3. Trial fields and response residuals}\label{lemma-3.-trial-fields-and-response-residuals}}

The translation term \(C^w\) uses the shifted-core structure to cancel the leading second-order source, while \(\mathbf J_{\rm dil}\) removes the remaining mixed-pair metric/dilation source. The quadratic exponential correction \(C^{\log}\) combines the nonlinear contacts so that the undifferentiated one contains the small density error in (S27). The phase lift \(\zeta_2\) matches the finite constant in the mixed second-order phase.

A regular trial field with exact first-order GP modes and all complementary second-order responses satisfies
\[
e_\rho\le C[L^{-1}+\sqrt qL^{39/2}+qL^{18}+q^2L^{30}]\longrightarrow0 .
\tag{S25}\label{eq:response-error}
\]

\emph{Construction and proof.} In Cartesian Nambu coordinates, restoring angular factors, set \(\Psi_0=(\psi_0,\psi_0^*)^T\), \(D_+=\partial_z/\sqrt\pi\), \(D_-=\partial_{\bar z}/\sqrt\pi\), \(z=x_1+ix_2\). Write signed legs as \(e_i=C_iD_i\Psi_0+\epsilon_i\), \(P_i=\mathsf H+\mu_i\), \(\mu_i=|k_i|^2/2-\omega_i\eta\), where \(\mathsf H\) is the zero-tangent Hessian. Differentiating the background and mode equations gives the exact cancellation
\[
\begin{aligned}
C^w_{ij}&=C_iD_ie_j+C_jD_je_i-C_iC_jD_iD_j\Psi_0,\\
P_{ij}C^w_{ij}+\mathbf T(e_i,e_j)
&=C_iC_j(k_i\cdot k_j)D_iD_j\Psi_0
+C_i(\mu_{ij}-\mu_j)D_i\epsilon_j\\
&\quad+C_j(\mu_{ij}-\mu_i)D_j\epsilon_i+\mathbf T(\epsilon_i,\epsilon_j).
\end{aligned}
\tag{S26}\label{eq:translation-cancellation}
\]

Here \(\mu_{ij}-\mu_i-\mu_j=(k_i\cdot k_j)I\), which vanishes for the perpendicular same-chirality pairs of the selected quartet. For mixed chirality, \(D_+D_-\Psi_0=(f_0^2-1)\Psi_0/(2\pi)=\mathsf H\mathbf J_{\rm dil}\), \(\mathbf J_{\rm dil}=-r\partial_r\Psi_0/(4\pi)\). Subtract \(C_iC_j(k_i\cdot k_j)\mathbf J_{\rm dil}\) and add \(C^{\log}_{ij}=(\psi_0\alpha_i\alpha_j,\psi_0^*\beta_i\beta_j)^T\), where \(\alpha_i=\epsilon_{i,u}/\psi_0\), \(\beta_i=\epsilon_{i,v}/\psi_0^*\). Its exact upper residual is
\[
\begin{aligned}
\frac{(P_{ij}C^{\log}_{ij}+\mathbf T(\epsilon_i,\epsilon_j))_u}{\psi_0}
={}&(k_i\cdot k_j)\alpha_i\alpha_j-\nabla_\perp\alpha_i\cdot\nabla_\perp\alpha_j\\
&+f_0^2(\alpha_i+\beta_i)(\alpha_j+\beta_j)
+\frac{E_{i,u}}{\psi_0}\alpha_j+\alpha_i\frac{E_{j,u}}{\psi_0},
\end{aligned}
\tag{S27}\label{eq:log-contact}
\]

where \(E_i=P_i\epsilon_i=-C_i\mu_iD_i\Psi_0\); conjugation gives the lower identity. The undifferentiated contact contains the small density error \(\alpha_i+\beta_i=\sqrt2\epsilon_{x,i}/f_0\), with its angular factor.

\emph{Trial fields.} Realize these coefficients on one moving graph: \(\mathbf X=(X,Y)\), \(w_i=-C_ia_i/\sqrt\pi\), \(A_X=(\partial_s\mathbf X)(\partial_s\mathbf X)^T\), \(\mathsf M=(I+A_X)^{1/2}\), \(\mathbf y=\mathsf M^{-1}(x_\perp-\mathbf X)\), and
\[
\psi_I=\psi_0(\mathbf y)e^{a(\mathbf y)+ib(\mathbf y)+i\zeta_2},\qquad
\Phi_{I,F}=\arg\mathbf y+b_{\rm BS}(\mathbf y)+\zeta_2 .
\tag{S28}\label{eq:inner-trials}
\]

Here \(a+ib\) is the first-order translation-subtracted GP relative field and \(b_{\rm BS}\) its circulation phase, including all legs and conjugates. The metric, shift and exponential give \(C^w-C_iC_j(k_i\cdot k_j)\mathbf J_{\rm dil}+C^{\log}+i\zeta_{ij}\eta\Psi_0\) for a mixed pair. In particular \(k_1\cdot k_{-4}=-2\lambda^2q^2/(1+\lambda^2)\).

The near-axis pole or logarithm and decay fix the second-order circulation phases:
\[
\Phi_s^{\rm BS}=\frac{iC_\lambda C_1q^2\kappa_s^2}{4\pi}K_2(\kappa_sqr)e^{-2i\vartheta},\quad
\Phi_b^{\rm BS}=\frac{iC_\lambda C_1q^2(\lambda^2-1)}{4\pi}K_0(\kappa_bqr),\quad
\Phi_0^{\rm BS}=0 .
\tag{S29}\label{eq:second-phases}
\]

They solve \((\partial_r^2+r^{-1}\partial_r-m_j^2/r^2-\kappa_j^2q^2)\Phi_j=0\), \(m_s=-2\), \(m_b=m_0=0\); complementary harmonics are conjugate. The inner sum phase is \(iC_\lambda C_1[\lambda^2q^2K_2(\lambda qr)+q^2K_2(qr)-2/r^2]e^{-2i\vartheta}/(4\pi)\), with sign from \(K_1'-K_1/z=-K_2\). The expansion \(z^2K_2(z)=2-z^2/2+O(z^4[1+|\ln z|])\), with \(\kappa_s^2=\lambda^2+1\), matches both pole and constant.

The unlifted mixed phase is \(iC_\lambda C_1q^2[\lambda^2K_0(\lambda qr)-K_0(qr)]/(4\pi)\). Choose
\[
\zeta_{\lambda,\mathsf C1}=\frac{iC_\lambda C_1q^2}{4\pi}
[\lambda^2\ln\lambda-(\lambda^2-1)\ln\kappa_b],\qquad
|\delta\phi_2|+r|\delta\phi_2'|\le Cq^4r^2[1+|\ln(qr)|],
\tag{S30}\label{eq:phase-join}
\]

with the conjugate sign in the complementary sector. This cancels the complete mixed constant and logarithm; \(\delta\phi_2=\Phi_{I,F}^{(2)}-\Phi_{\rm BS}^{(2)}\) on \(r<3R_q\). The zero static phase retains its GP density response. Coefficients are bounded on \(I\); first-field Bessel decay rates at least \(a-1/40\) have pairwise sums \(>19/10>\max_I\kappa_j\), controlling growing Bessel weights over the extended sources.

Choose an Eulerian radial cutoff \(\chi=1\) below \(2R_q\), \(\chi=0\) above \(3R_q\). With the exact linear density \(n_1\) and phase correction \(\delta\phi_1=\phi_{1,\rm GP}-\phi_{1,\rm BS}\), define
\[
\begin{aligned}
n_G&=\chi|\psi_I|^2+(1-\chi)(f_0^2+n_1),\\
\Phi_G&=\chi\arg\psi_I+(1-\chi)(\Phi_{\rm BS}+\delta\phi_1),\\
\Phi_F&=\chi\Phi_{I,F}+(1-\chi)\Phi_{\rm BS}.
\end{aligned}
\tag{S31}\label{eq:global-trials}
\]

The phase differences are single valued; \((\sqrt{n_G}e^{i\Phi_G},\sqrt{n_G}e^{-i\Phi_G})^T\) is core-regular with exact first modes. Extract second coefficients \(C_{ij}\) without factorials and set \(\rho_{ij}=P_{ij}C_{ij}+\mathbf T_{ij}\). Equation (S17) covers \((1,2),(-3,-4),(1,-4),(2,-3),(2,-4),(1,-3)\) and their conjugates.

\emph{Residual bounds.} Equations (S21), (S26)--(S27) give inner forces \(O(q^4L^5)\), the remaining mixed dilation force \(O(q^4L/r^2)\), and a lift contribution. For the exterior, the exact second Bernoulli and continuity residuals are
\[
\begin{gathered}
\mathcal B_{ij}=(\partial_t+\nabla\vartheta\cdot\nabla)\Phi_{ij}^{\rm BS}
+\nabla\phi_i\cdot\nabla\phi_j+Q_2(n_i,n_j),\quad
\mathcal C_{ij}=\nabla\cdot\mathcal J_{ij},\\
\mathcal J_{ij}=(f_0^2-1)\nabla\Phi_{ij}^{\rm BS}+n_i\nabla\phi_j+n_j\nabla\phi_i,\\
Q_2=\frac1{4f_0^2}\nabla\cdot[f_0^2\nabla(h_ih_j)]-\tfrac14\nabla h_i\cdot\nabla h_j,\quad
h_i=n_i/f_0^2,\quad n_i=\sqrt2f_0x_i,\quad\phi_i=y_i/(i\sqrt2f_0).
\end{gathered}
\tag{S32}\label{eq:exterior-residuals}
\]

All derivatives include tangent labels. The Cartesian force is \((\sqrt2f_0\mathcal B,-i\mathcal C/(\sqrt2f_0))\), retaining contacts even where quadratic density is zero. With \(\Omega=C_\omega q^2L\), (S22) gives \(|\mathcal B|\le C(r^{-4}+\Omega r^{-2})\), \(|\mathcal J|\le C(r^{-5}+\Omega r^{-3})\). Pairing the divergence before estimating uses the phase gradient form in (S18); only the cutoff divergence costs \(q^{-1}\). In the \(m=-2\) channel its centrifugal form and \(\mathsf B_0\ge1/(4r^2)\) also control the angular phase derivative.

At the join, \([P,\chi]\delta C=-\nabla\chi\cdot\nabla\delta C-(\Delta\chi)\delta C/2\). The first fields agree, cancelling Cartesian contact differences. The product rule and (S22), (S30) give, at \(R=R_q\),
\[
\begin{aligned}
|\delta n_2|+R|\delta n_2'|&\le C(R^{-4}+q^2L^{5/2}R^{-2}+q^4L^5R^{-2}),\\
|\delta\Phi_2|+R|\delta\Phi_2'|&\le Cq^2[L^{-12}(1+\ln L)+\sqrt qL^{27/2}+q^2L^{24}].
\end{aligned}
\tag{S33}\label{eq:join-errors}
\]

Density derivatives use second derivatives of the modes and translation; the phase retains \(f_0^{-1}\). The normalized bounds are:

\par\addvspace{\medskipamount}
\noindent\begin{minipage}{\textwidth}
\textbf{Table S4.} Response residuals, uniform on \(I\).
\par\smallskip\centering
\begin{tabular}{@{}ll@{}}
\toprule
Contribution & Bound for \(q^{-2}\|\rho\|_{P^{-1}}\), apart from \(C\) \\
\midrule
Inner force and lift & \(qL^5+L^{-1}+L^{-5}\) \\
Exterior & \(q(L^{18}+L^7)+q^2(L^{30}+L^{19})\) \\
Density commutator & \(q^3L^{30}+q^5L^{23}\) \\
Phase commutator & \(L^{-6}(1+\ln L)+\sqrt qL^{39/2}+q^2L^{30}\) \\
\bottomrule
\end{tabular}
\end{minipage}\par\medskip

The exterior norm is \(\le C[R^{-3}+\Omega R^{-1}+q^{-1}R^{-5}+\Omega q^{-1}R^{-3}]\); the phase commutator costs \((qR)^{-1}=L^6\). Setting \(R=R_q\) gives the table and (S25) for every complementary pair. \(\square\)

\hypertarget{lemma-4.-stationarity-and-the-gpfluid-action}{%
\subsection{Lemma 4. Stationarity and the GP--fluid action}\label{lemma-4.-stationarity-and-the-gpfluid-action}}

For the preceding trial,
\[
|\Gamma_{S,G}-\Gamma_{S,T}|\le3q^4e_\rho^2,\qquad
e_A\le C[L^{-7}+\sqrt qL^P+q^2L^P]\longrightarrow0 .
\tag{S34}\label{eq:action-errors}
\]

\emph{Proof.} For the physical second-order field \(Z^{(2)}\), write the quartic action
\[
\mathcal A_4[Z^{(2)}]=D_4-\tfrac12\langle Z^{(2)},\mathscr PZ^{(2)}\rangle
-\langle Z^{(2)},\mathbf T_2\rangle,\qquad
\mathcal A_4[Z_*^{(2)}]-\mathcal A_4[Z^{(2)}]
=\tfrac12\langle\rho_2,\mathscr P^{-1}\rho_2\rangle .
\tag{S35}\label{eq:stationary-action}
\]

Here \(Z_*^{(2)}=-\mathscr P^{-1}\mathbf T_2\), with direct term \(D_4\). For each partition \(p|p^c\) of \(M=a_1a_2a_3^*a_4^*\), the sector residual is \(A_p\rho_p+\overline{A_{p^c}}\mathsf C\rho_{p^c}\). Resonance gives \(P_p\mathsf C=\mathsf CP_{p^c}\); conjugate sectors cancel \(1/2\), producing \(M\) once:
\[
\begin{aligned}
\Gamma_{S,G}-\Gamma_{S,T}
={}&\langle\rho_{3,4},P_{1,2}^{-1}\rho_{1,2}\rangle
+\langle\rho_{3,-2},P_{1,-4}^{-1}\rho_{1,-4}\rangle\\
&+\langle\rho_{3,-1},P_{2,-4}^{-1}\rho_{2,-4}\rangle .
\end{aligned}
\tag{S36}\label{eq:complementary-polarization}
\]

The bras are complementary particle--hole residuals; Cauchy--Schwarz in the positive response norm gives three. Unit Krein norm fixes the quadratic canonical Berry term, and variation in \(a_4^*\) gives (S13): \(g_{\rm GP}=-\Gamma_{S,G}\).

For the full-field action, Green's identity leaves background/fourth-, first/third-, second/second-order and temporal boundary terms. Exact background/first equations, regular core jets and one global trial cancel the interior terms. At fixed \(q>0\), first and nonzero-transfer second fields decay; selected third transfers are opposite to external legs, required third/neutral-fourth jets are bounded, and \(f_0'=O(R_{\rm IR}^{-3})\). Thus outer terms vanish as \(R_{\rm IR}\to\infty\); resonant temporal endpoints agree. Taking only \(R_{\rm IR}\sim q^{-1}\) would permit an \(O(q^4)\) first/third boundary term.

\emph{Inner region.} We compare the complete density-phase Lagrangian \(\mathscr L_{\rm GP}=-n\partial_t\Phi-n|\nabla\Phi|^2/2-(n-1)^2/2-|\nabla n|^2/(8n)\). Use the exact coordinate identity with \(J_A=\det\mathsf M\), \(\phi=\psi_0(\mathbf y)e^{a+ib+i\zeta_2}\), \(B_\mu=\mathsf M^{-1}\partial_\mu\mathsf M\), \(U_\mu=\mathsf M^{-1}\partial_\mu\mathbf X\), and \(\mathcal D_\alpha=\partial_\alpha-B_\alpha\mathbf y\cdot\nabla_\perp\). On \(|\mathbf y|<R\),
\[
\begin{aligned}
E_I&=\int J_A\left\{e_\perp(\phi)+\tfrac12\sum_\alpha|\mathcal D_\alpha\phi|^2
-\sum_\alpha\operatorname{Re}[(U_\alpha\cdot\nabla_\perp\phi)^*\mathcal D_\alpha\phi]\right\}d^2y,\\
\mathscr B_I&=-J_A f_0^2e^{2a}
\left[\partial_t(b+\zeta_2)-(U_t+B_t\mathbf y)\cdot\nabla_\perp(\vartheta+b)\right],\\
2e_\perp(\phi)&=e^{2a}\{|\nabla f_0+f_0\nabla a|^2+f_0^2|\nabla(\vartheta+b)|^2\}
+(f_0^2e^{2a}-1)^2 .
\end{aligned}
\tag{S37}\label{eq:exact-inner-action}
\]

Here \(\mathsf M^{-1}(I+A_X)\mathsf M^{-T}=I\) and \(\nabla_\perp\zeta_2=0\). Scaling \((\mathbf X,a,b,\zeta_2)\mapsto(\epsilon\mathbf X,\epsilon a,\epsilon b,\epsilon^2\zeta_2)\) and extracting \(\epsilon^4\) retains every quartic. Write \(n=n_0+\sum_{j\ge1}\epsilon^jn_j\), \(\Phi=\vartheta+\sum_{j\ge1}\epsilon^j\phi_j\), \(v_j=\nabla\phi_j\), \(v_0=\nabla\vartheta\). Subscripts mean homogeneous amplitude degree, including all monomials without factorials; G and BS denote GP and circulation fields. GP has \(n_0=f_0^2\), fluid \(n\equiv1\). Constant slope with \(a=b=\zeta_2=0\) gives exactly \(\tau_{\rm GP}(R)J_A\); its selected quartic vanishes by (S1).

Use \(a=O(q^2L^{5/2}/r)\), \(\nabla a=O(q^2L^{5/2}/r^2)\), \(b=O(q^2rL^{5/2})\), \(\nabla b=O(q^2L^{5/2})\), with regular core estimates at zero; tangent/time derivatives cost \(q\)/\(q^2L\). The frame \(B_s=O(q^3)\) starts at degree two, \(U_s\) at degree one. With \(J_2=\operatorname{tr}A_X/2\), the largest spatial quartics are \(J_2|\nabla b|^2\) and products of \(\partial_sb\) with its degree-three frame correction or \(J_2\partial_sb\), integrating to \(q^6L^5R^2\), \(q^8L^5R^4\). Others gain density or inverse-radius factors; undifferentiated \(b\) is absent and the pure frame cross term is odd. The fourth antisymmetric Berry frame is \([\partial_tA_X,A_X]/8=O(q^6L)\); its symmetric part contracts to zero angularly with \(\nabla\vartheta\). The largest dressed integral is \(q^6L^{7/2}R^2\). The included lift terms have \(\nabla_s\zeta_2=O(q^3)\), \(\partial_t\zeta_2=O(q^4L)\).

The integrated \(\operatorname{Pol}_4\) bounds follow, with \(\mathcal H=R^{-6}+\Omega R^{-4}+\Omega^2R^{-2}\), \(\Omega=C_\omega q^2L\). Each \(L^P\) multiplies a term retaining a positive power of \(q\) after division by \(q^4\).

\par\addvspace{\medskipamount}
\noindent\begin{minipage}{\textwidth}
\textbf{Table S5.} Integrated action bounds after full-area subtraction.
\par\smallskip\centering
% Native p columns and e-TeX dimensions avoid array/calc dependencies.
\begin{tabular}{@{}
  p{\dimexpr 0.50\linewidth-\tabcolsep\relax}
  p{\dimexpr 0.50\linewidth-\tabcolsep\relax}@{}}
\toprule
\raggedright Region & \raggedright Integrated quartic bound \tabularnewline
\midrule
\raggedright GP inner, after full-area subtraction, including lift and Berry terms & \raggedright \(C[q^6L^5R^2+q^8L^5R^4+q^6L^P]\) \tabularnewline
\raggedright Fluid inner, after full-area subtraction & \raggedright \(C[(q^6R^2+q^8R^4)(1+\ln^2(qR))+q^6LR^2(1+|\ln(qR)|)+q^6L^P]\) \tabularnewline
\raggedright Annulus & \raggedright \(CL^P[\varepsilon_qR^{-4}+\mathcal H+R^{-8}+\Omega R^{-6}+\Omega^2R^{-4}]\) \tabularnewline
\raggedright Moving interface & \raggedright \(CL^P\{\varepsilon_q[R^{-4}+(\Omega+q^2)R^{-2}+q^2\Omega]+\mathcal H+q^2R^{-4}+q^2\Omega R^{-2}+q^4R^{-2}\}\) \tabularnewline
\raggedright Exterior & \raggedright \(CL^P[\varepsilon_qR^{-4}+\mathcal H+qR^{-5}+q^2R^{-4}]\) \tabularnewline
\bottomrule
\end{tabular}
\end{minipage}\par\medskip

On the annulus, \(|\partial_r^an_j|\le CL^P(r^{-2}+\Omega)r^{-j-a}\), \(a\le4-j\); phase differences are \(O(\varepsilon_qr^{-j})\), gaining an inverse radius per derivative. The complete term linear in fourth density is
\[
-[|\nabla\vartheta|^2/2+f_0^2-1+Q_{\rm qp}(f_0^2)]n_4
-\nabla\cdot\frac{n_4\nabla f_0^2}{4f_0^2},
\qquad Q_{\rm qp}(n)=-\frac{\Delta\sqrt n}{2\sqrt n}.
\tag{S38}\label{eq:fourth-density}
\]

The background equation cancels the bracket; the circle flux is \(O(L^P[R^{-8}+\Omega R^{-6}])\), requiring no fourth derivative of \(n_4\). Pressure \(n_1n_3+n_2^2/2\) integrates to \(CL^P\mathcal H\); remaining quantum pressure gains two inverse radii; kinetic and Berry differences give \(CL^P(\varepsilon_qR^{-4}+\mathcal H)\).

\emph{Moving interface.} For the moving interface use its exact ellipse \(|(I+\epsilon^2A_X)^{-1/2}(x-\epsilon\mathbf X)|<R\), with radial boundary \(R+h(\epsilon,\vartheta)\). Define
\[
M_jF=[\epsilon^j]\int d\vartheta\int_R^{R+h(\epsilon,\vartheta)}rF(r,\vartheta)\,dr,\qquad
D_kF=\sum_{a\le k}R^a\sup|\partial_r^aF|.
\]

The coefficients satisfy \(h_1=O(1)\), \(h_2=O(R^{-1}+q^2R)\), \(h_3=O(q^2)\), \(h_4=O(R^{-3}+q^2R^{-1}+q^4R)\). Taylor expansion of this radial primitive gives
\[
\begin{aligned}
|M_1F|&\le CRD_0F,& |M_2F|&\le C(D_1F+q^2R^2D_0F),\\
|M_3F|&\le C(R^{-1}D_2F+q^2RD_1F),&
|M_4F|&\le C(R^{-2}D_3F+q^2D_2F+q^4R^2D_1F).
\end{aligned}
\tag{S39}\label{eq:moving-boundary}
\]

Applied to \(\sum_{j=1}^4M_j[\Delta\mathscr L_{4-j}]\), these give the interface row using three radial derivatives of lower-degree Lagrangians, including their Berry frequencies.

The circulation current has at most two slope factors. Nonzero second/third transfers give \(|v_2|\le Cr^{-3}\), \(|v_3|\le Cr^{-4}\); the neutral fourth uses derivatives of \(G_0=-\ln r/(2\pi)\), giving \(|v_4|\le C(r^{-5}+qr^{-4}+q^2r^{-3})\). Its common unit-density \(v_0\cdot v_4\) cancels between actions; the density-weighted remainder integrates to \(C(R^{-6}+qR^{-5}+q^2R^{-4})\). Propagate (S22) on intervals of size \(r\) for \(qr\le1\), \(q^{-1}\) thereafter:
\[
|\nabla_\perp^h(v_{1,G}-v_{1,\rm BS})|\le C_h\varepsilon_qr^{-2-h}\ (h\le3),\qquad
|n_1|\le C[r^{-3}+q^2L^{3/2}\ln(2r)/r].
\]

The phase retains \(f_0^{-1}\). These bounds give the exterior row, including \(-n_1\partial_t\phi_3\) and quartic quantum pressure. Resonance removes the genuine fourth time derivative; lower-frequency Berry products remain.

At \(R=R_q\), the largest spatial ratios to \(q^4\) are \(L^{-7},L^{-19}\), and the dressed Berry ratio is \(L^{-17/2}\). Fluid inner ratios are \(\le C[L^{-12}(1+\ln^2L)+L^{-11}(1+\ln L)]\). Others retain positive powers of \(q\), e.g.~\(q^{-4}\varepsilon_qR^{-4}=\sqrt qL^{51/2}+q^2L^{36}\), proving (S34). \(\square\)

\hypertarget{lemma-5.-the-finite-fluid-vertex}{%
\subsection{Lemma 5. The finite fluid vertex}\label{lemma-5.-the-finite-fluid-vertex}}

The complete fluid action satisfies
\[
e_F\le C[\sqrt qL+L^{-24}(1+\ln^2L)]\longrightarrow0 .
\tag{S40}\label{eq:fluid-error}
\]

\emph{Proof.} We compare the energies first, then the Berry forms, and finally match the core regularization and canonical normalization. The trial and circulation fields share graph, meridional circulation and tangent periods. Their difference is \(\nabla\xi_{\rm pot}\), with single-valued \(\xi_{\rm pot}\) supported below \(3R_q\) and starting at degree two. On a common tube of radius \(\epsilon\), singular circulation is tangential and the remaining normal velocity \(O(1+|\ln\epsilon|)\). After common-area subtraction, integration by parts gives
\[
E_F^{\rm ren}-E_{\rm BS}^{\rm ren}=\tfrac12\int|\nabla\xi_{\rm pot}|^2,\qquad
\operatorname{Pol}_4(E_F^{\rm ren}-E_{\rm BS}^{\rm ren})
=\tfrac12\operatorname{Pol}_4\int|\nabla\delta\phi_2|^2 .
\tag{S41}\label{eq:fluid-orthogonality}
\]

Equation (S30), including cutoff derivatives, yields \(\int|\nabla\delta\phi_2|^2\le Cq^8R_q^4[1+\ln^2(qR_q)]\). The origin is integrable, and moving-domain corrections to a square starting at degree four begin at degree five.

To commute four amplitude derivatives with the core limit, project source patches onto the moving tangent plane and subtract \(\Gamma d\vartheta/(2\pi)\) in its orthonormal frame. Displacement and orientation remainders are \(O(t^2)\), \(O(t)\); the differentiated \(|\mathcal R|^{-3}\) velocity kernel is bounded by
\[
\int_0^a\frac{t^{2j+1}}{(r^2+t^2)^{(3+j)/2}}\,dt .
\]

This is logarithmic for \(j=1\) and bounded for \(j\ge2\); orientation terms are also logarithmic. Bounded frame/patch derivatives then give omitted flux \(O(\epsilon[1+|\ln\epsilon|]^P)\) and square integral \(O(\epsilon^2[1+|\ln\epsilon|]^P)\), through four amplitude derivatives at fixed \(q\).

For the Berry form keep the infrared boundary fixed. On the normal circle, \(\delta_U\Phi=-\Gamma U_\perp\cdot e_\vartheta/(2\pi\epsilon)+O(|\ln\epsilon|)\). Reynolds differentiation gives the normal oriented determinant; true area cancels the graph-normal projection determinant:
\[
d\Theta_F=-\Gamma\int ds\,\delta X_1\wedge\delta X_2,\qquad
\Theta_{\rm can}=\tfrac\Gamma2\int ds\,(X_2\delta X_1-X_1\delta X_2).
\tag{S42}\label{eq:fluid-berry}
\]

Boundary errors through the required amplitude derivatives are \(O(\epsilon[1+|\ln\epsilon|]^P)\). Radial homotopy gives \(\Theta_\epsilon-\Theta_{\rm can}=dF_\epsilon+\mathcal R_\epsilon\), with
\[
\begin{aligned}
F_\epsilon(\alpha)&=\int_0^1(\Theta_\epsilon-\Theta_{\rm can})_{\sigma\alpha}[\alpha]\,d\sigma,\\
\mathcal R_\epsilon(\alpha)[U]&=\int_0^1\sigma\,d(\Theta_\epsilon-\Theta_{\rm can})_{\sigma\alpha}(\alpha,U)\,d\sigma\longrightarrow0.
\end{aligned}
\]

Under the linear mode evolution, the selected monomial carries phase \(e^{-i(\omega_1+\omega_2-\omega_3-\omega_4)t}=1\) at exact resonance. Thus the total derivative \(dF_\epsilon/dt\) has no selected quartic coefficient. The canonical quadratic form has no resonant quartic Berry term.

It remains to compare the circulation energy with S1's regulated Hamiltonian. Mollify \(G_4(x)=1/(4\pi^2|x|^2)\) by a normalized radial \(\varrho_\epsilon\) supported within \(\epsilon\). Outside its tube harmonic velocity is unchanged. Inside, the \(O(\epsilon^{-1})\) flat field's cross term with the logarithmic curvature remainder integrates to \(O(\epsilon|\ln\epsilon|)\); surviving local energy is true area. The mollified-minus-softened kernel obeys
\[
\begin{gathered}
D_\epsilon=G_4*(\varrho_\epsilon*\varrho_\epsilon)-[4\pi^2(|x|^2+\epsilon^2)]^{-1},\\
|\nabla^jD_\epsilon(x)|\le\frac{C_j\epsilon^2}{(|x|+\epsilon)^{4+j}}\quad(0\le j\le5),\qquad
\int_{\mathbb R^2}|t|^{j+1}|\nabla^jD_\epsilon(t)|\,d^2t\le C_j\epsilon .
\end{gathered}
\tag{S43}\label{eq:regulator-moments}
\]

Subtracting exact affine-plane energy makes orientation and chord remainders gain one and two separation powers. The moments control four amplitude derivatives, including the noncompact softened tail. Both regulator changes are \(\delta\tau\,\operatorname{Area}(\mathbf X)+o(1)\), with zero selected quartic. Thus (S8) includes the full circulation energy, including first/third and background/fourth terms.

The canonical Berry form makes the resonant action vertex minus the corresponding energy vertex. Finally \(w_i=-C_ia_i/\sqrt\pi\) converts the vertex by \(\prod_iC_i=1+O(\sqrt qL)\); four minus signs cancel. Combining this calibration with (S41) proves (S40).

For rational \(\lambda^2=M/N\), use the reciprocal lattice generated by \((\lambda q/(M+N),0)\), \((0,q/(M+N))\). Every proper nonempty selected signed subset is a leg, pair transfer or triple opposite to a leg, of magnitude \(\ge24q/(5\sqrt{74})\). Only three quadratic responses are inverted; prescribed higher fields retain the neutral fourth algebraic tail. Its Green-function constant changes only background energy, since circulation is fixed and the torus means of \(\partial_s\mathbf X\), \(\det\partial_s\mathbf X\) vanish. At fixed \(q,\lambda\), the common core limit through four amplitude derivatives leaves only local area terms; per-area bounds depend on physical transfers, independently of cell size. Continuity of the simple modes in the elliptic \(H^2\) graph norm and of positive response resolvents extends the bound to all \(\lambda\in I\). \(\square\)

\hypertarget{completion-of-the-theorem}{%
\subsection{Completion of the theorem}\label{completion-of-the-theorem}}

Since \(g_{\rm GP}=-\Gamma_{S,G}\), the triangle inequality, (S25), (S34) and (S40) give (S16) on one common small-\(q\) interval. The strict endpoint bounds \(0.0208<F_\perp(49/25,1)<0.0210\), \(-0.0094<F_\perp(9/4,1)<-0.0092\) follow from \(\ln x=2\sum_{j=0}^{n-1}z^{2j+1}/(2j+1)+R_n\), \(z=(x-1)/(x+1)\), with \(|R_n|\le2|z|^{2n+1}/[(2n+1)(1-z^2)]\); twelve terms give total substitution error below \(6\times10^{-8}\). Since \(g_{\rm GP}\) is real and continuous, the opposite endpoint signs give a zero in the open interval for sufficiently small \(q\).

To identify its limit, put \(x=\lambda^2\). Differentiating (S8) gives
\[
\begin{gathered}
f'(\lambda)=\frac{\lambda}{2\pi}G(x),\quad
G(x)=(x-1)\ln\frac{x-1}{x}+\ln\frac{x+1}{x}+\frac{x}{x+1}-\frac{8x}{(x+1)^3},\\
G''(x)=\frac{x^4-40x^3+110x^2-40x+1}{x(x-1)(x+1)^5}>0
\quad (1<x\le9/4).
\end{gathered}
\tag{S44}\label{eq:fluid-monotonicity}
\]

For the numerator, division by \(x^2\) gives \(y^2-40y+108>0\) for \(2<y=x+x^{-1}\le97/36\). Convexity bounds \(G\) on \([49/25,9/4]\) above by its endpoints. The elementary logarithmic bounds \(\ln(24/49)<-7/10\), \(\ln(74/49)<5/12\), \(\ln(5/9)<-7/12\), \(\ln(13/9)<3/8\) give both endpoints below \(-9/50\). Hence \(f'<-63/(500\pi)<-1/25\) on \(I\), so the fluid zero is unique there.

To exclude an earlier zero, \(G(1^+)=\ln2-1/2>0\), whereas \(G(49/25)<0\). Strict convexity gives a unique crossing from positive to negative on \((1,49/25)\): at any zero its derivative is negative by comparison with the secant to \(49/25\). Thus \(f\) has only a maximum on \([1,7/5]\); its minimum is at an endpoint, both positive since \(f(1)=(2\ln2-1)/(8\pi)>0\) and \(f(7/5)>0\). Consequently \(\lambda_*\) is the first fluid zero for \(\lambda>1\).

With \(c_f=\min_I(-f')>0\) and \(\epsilon(q)=\sup_I|q^{-4}g_{\rm GP}-f|\), every microscopic zero in \(I\) satisfies
\[
|\lambda_q-\lambda_*|\le\epsilon(q)/c_f\longrightarrow0.
\tag{S45}\label{eq:root-convergence}
\]

This uses uniform convergence without differentiating \(g_{\rm GP}\); fixed-\(q\) uniqueness and a sharp root-shift law are not asserted. Diagonal shifts and higher-order conversion remain allowed.

\hypertarget{s5.-numerical-methods-and-sensitivity-controls}{%
\section{S5. Numerical methods and sensitivity controls}\label{s5.-numerical-methods-and-sensitivity-controls}}

\textbf{Reflected-family coefficients.} The modes, sources and signed contraction are defined in Sec. S2. Write \(r=r_\perp\). Factor the background as \(f_0=rg_0\), impose \(g_0'(0)=0\) and \(f_0(R)=1-1/(4R^2)-9/(32R^4)\). For angular label \(m\), factor \(u=r^{|m+1|}U\), \(v=r^{|m-1|}V\), with \(U'(0)=V'(0)=0\) and \(u(R)=v(R)=0\); singular derivative terms are handled analytically at the included origin. Kelvin modes solve \(L_{-1}(K)Z=\omega Z\), with integral-state unit positive Krein norm and real phase \(u(0)>0\). Paired magnitudes share radial functions.

The contraction exchanges outgoing labels, taking the low outgoing leg as conjugated input and the high outgoing leg as receiver. Table S1 then specifies all three responses from \(Z_{\lambda q},Z_q,\mathsf C Z_q\); both components and conjugates enter, without broadening. Adaptive collocation starts on quadratic meshes, with 16384 nodes per field and background tolerance \(10^{-12}\). Twelve-point Gauss quadrature on the saved cubic interpolants' union mesh integrates projection polynomials of degree at most 21, up to floating-point evaluation, without removing field discretization error.

Finite-volume cells have \(r_j=(j+1/2)h\), weights \(2\pi r_jh\), zero origin-face flux and the same outer data. The estimate \([4g(h_f)-g(2h_f)]/3\) assumes leading \(h^2\) error. In Fig. 2(b), raw and estimated coefficients bracket a sign change in {[}1.35,1.40{]}; \(h=1/512\) gives {[}1.40,\(\lambda_*\){]}.

\par\addvspace{\medskipamount}
\noindent\begin{minipage}{\textwidth}
\textbf{Table S6.} Finite-volume controls at \(q=0.1,R=256,h_f=1/1024\), in \(10^{-6}\).
\par\smallskip\centering
\begin{tabular}{@{}rrr@{}}
\toprule
\(\lambda\) & Finest raw \(g\) & Empirical \(h^2\) estimate \\
\midrule
1.35 & +0.217600552 & +0.163733085 \\
1.40 & -0.052014368 & -0.105977732 \\
\(\lambda_*\) & -0.428333203 & -0.482430689 \\
\bottomrule
\end{tabular}
\end{minipage}\par\medskip

\par\addvspace{\medskipamount}
\noindent\begin{minipage}{\textwidth}
\textbf{Table S7.} Predeclared collocation controls, with completed/requested counts. The radii give \(qR=12.8,25.6\); tolerances apply to modes and responses.
\par\smallskip\centering
\begin{tabular}{@{}rlllr@{}}
\toprule
\(q\) & \(\lambda\) & \(R\) & Tolerances & Count \\
\midrule
0.1 & 1.35, 1.40 & 128, 256 & \(10^{-10},10^{-11}\) & 8/8 \\
0.05 & 1.40, 1.45, 1.50 & 256, 512 & \(10^{-10},10^{-11}\) & 12/12 \\
0.025 & 1.40, 1.45, 1.50 & 512, 1024 & \(10^{-10},10^{-11},10^{-12}\) & 17/18 \\
\bottomrule
\end{tabular}
\end{minipage}\par\medskip

\par\addvspace{\medskipamount}
\noindent\begin{minipage}{\textwidth}
\textbf{Table S8.} \(q=0.1\) controls, in \(10^{-6}\): extrema over Table S7 settings.
\par\smallskip\centering
\begin{tabular}{@{}rrr@{}}
\toprule
\(\lambda\) & Minimum \(10^6g_{\rm GP}^{(R)}\) & Maximum \(10^6g_{\rm GP}^{(R)}\) \\
\midrule
1.35 & +0.163729667 & +0.163768662 \\
1.40 & -0.105978984 & -0.105967913 \\
\bottomrule
\end{tabular}
\end{minipage}\par\medskip

\par\addvspace{\medskipamount}
\noindent\begin{minipage}{\textwidth}
\textbf{Table S9.} Longer-wave controls, in \(10^{-9}\): extrema over successful Table S7 settings. Digits describe sensitivity, without certified accuracy. The failed \(q=0.025,\lambda=1.50\) request is detailed below.
\par\smallskip\centering
\begin{tabular}{@{}rrrr@{}}
\toprule
\(q\) & \(\lambda\) & Minimum \(10^9g_{\rm GP}^{(R)}\) & Maximum \(10^9g_{\rm GP}^{(R)}\) \\
\midrule
0.05 & 1.40 & +8.929591393 & +8.936700706 \\
0.05 & 1.45 & -7.001117318 & -6.994484138 \\
0.05 & 1.50 & -22.874288172 & -22.867846741 \\
0.025 & 1.40 & +0.987295815 & +1.007230500 \\
0.025 & 1.45 & +0.025291498 & +0.045052594 \\
0.025 & 1.50 & -0.932418087 & -0.912821069 \\
\bottomrule
\end{tabular}
\end{minipage}\par\medskip

All 37 successes retain sampled endpoint signs. Near zero, \(q=0.025,\lambda=1.45\), the full spread is 78.13\%; the last tolerance step changes the coefficient by up to 6.68\%. At \(10^{-12}\), changing \(R=512\) to 1024 changes \(4.50525936\times10^{-11}\) to \(4.48519832\times10^{-11}\), or 0.447\%. These are sensitivities, not error bars. At \(q=0.025,\lambda=1.50,R=1024\), the \(10^{-12}\) high-mode solve exceeded the refinement ceiling (maximum RMS residual \(2.18\times10^{-11}\)), producing no coefficient.

Independent reconstruction from all 37 saved field sets checks GP sources, equations, boundaries, norms and reflection phase without re-solving. Direct/mediated contractions agree within \(5.56\times10^{-17}\) despite individual terms reaching \(2.34\times10^{10}\) times the final coefficient. The readback thus tests saved-field and contraction consistency under strong cancellation; it does not certify lowest-branch identity, spectral distance or coefficient error. Figure 2(a) includes every successful control, with overlapping symbols.

\textbf{Coherent evolution.} Write \(\psi=e^{i\vartheta}\Phi(r,\phi_1,\phi_2,\phi_3)\), \(\phi_i=k_i\cdot s-\vartheta\). Fourier index \(j\in\mathbb Z^3\) has angular order \(1-\sum j_i\) and momentum \(\sum j_i k_i\), avoiding momentum rounding. All \(8^3\) phase bins evolve; receiver index \((1,1,-1)\) is distinct from other cubic indices. Symmetric splitting uses radial Crank--Nicolson, exact tangent phases and a pointwise nonlinear phase; opposite \(f_0^2\) shifts preserve the stationary profile without altering GP evolution. All runs use \(R=64\), Table S2 spacings, and \(\Delta t=0.005\) except two half-step controls. Six input-amplitude trajectories end at \(T=5\) for \(q=0.4\), and 0.5/0.25 for \(q=0.2,A=0.01/0.005\); the controls end at 1 and 0.25, respectively.

\textbf{GP parity spectra.} The actual tangent torus has reciprocal lattice \(\mathbb Z\mathbf u+\mathbb Z\mathbf v\), twice the primitive pump-cell area, with no normal-rotation boundary identification. Expand \(\psi=e^{i\vartheta}\sum_{jl}F_{jl}(r)e^{i(j\phi_-+l\phi_+)}\), \(\phi_\pm=p_\pm\cdot s-\vartheta-\Omega t\). Real reflection-symmetric backgrounds satisfy
\[
\begin{gathered}
0=[T_{1-j-l}+K_{jl}^2/2-\Omega(j+l)-1]F_{jl}+[|F|^2F]_{jl},\\
T_m=-\tfrac12(\partial_r^2+r^{-1}\partial_r-m^2/r^2),\quad K_{jl}^2=q^2(j^2+l^2+2djl),\\
\sum_{jl}(j+l)\int_0^R2\pi r|F_{jl}|^2dr=2N.
\end{gathered}
\tag{S46}\label{eq:parity-background}
\]

Use \(q=0.2,R=64,h=1/128\). Background labels run from \(-3\) to 3, probe labels from \(-3.5\) to 3.5, with 16 phase points per direction. The probe particle/hole angular orders are \(2-j-l,-j-l\), with drift \(-\Omega(j+l)\); both bare probes have the same \(-2\Omega\) shift. Only the zero background harmonic has reservoir boundary data; other harmonics and probe components vanish at \(R\), using the last half-cell distance. The six background and twelve probe weighted residuals are \(1.41\)--\(1.49\times10^{-10}\) and \(2.39\)--\(2.72\times10^{-12}\); reflection leakage vanishes and unit-Krein discrepancies are below \(3\times10^{-14}\).

\hypertarget{s6.-angular-conversion-and-the-soft-momentum-relation}{%
\section{S6. Angular conversion and the soft-momentum relation}\label{s6.-angular-conversion-and-the-soft-momentum-relation}}

For reflected incoming momenta, set \(a=|k_1/q|^2\), \(b=|k_2/q|^2\), \(c=k_1\cdot k_2/q^2\), and \(D=a+b+2c\). Reflection about \(k_1+k_2\) preserves paired magnitudes and exact isotropic resonance. For \(a\ne b\), \(c^2<ab\), substitution in the finite-vertex generator of S1 gives
\[
\begin{aligned}
\mathcal F_{\rm refl}(a,b,c)
&=\tfrac14[(a-b)^2\ln|a-b|-a(a+2c)\ln a-b(b+2c)\ln b]\\
&\quad+\tfrac12(a+c)(b+c)\ln D+\tfrac14ab
-\frac{2(ab-c^2)(a+c)(b+c)}{D^2},\\
\frac{V_{\rm can}}{q^4}
&=\frac{\mathcal F_{\rm refl}(\lambda^2,1,\lambda\cos\theta)-\mathcal L\lambda^2\cos^2\theta}{2\pi}.
\end{aligned}
\tag{S47}\label{eq:angular-coefficient}
\]

Here \(\mathcal L\) is fixed by \(\omega_{\rm BS}(k)=\tfrac12k^2[\mathcal L(q)-\ln(k/q)-\tfrac12]\). Perpendicular incidence removes the tension term and recovers \(f(\lambda)\); angular zeros concern the fluid theory.

For the soft limit {[}S10{]}, keep \(a_c>0\), incoming \(\mathbf k,\mathbf l\), total \(\mathbf Q=\mathbf k+\mathbf l\), and outgoing \(\mathbf Q-\mathbf p,\mathbf p\). Hard momenta occupy a compact set of scale \(k_h\), away from additional zero or forward transfers. With the transforms \(F_n\) of S1, integration by parts gives
\[
\begin{gathered}
h(\mathbf k)=K_E[k^2F_1(\mathbf k)+2F_2(\mathbf k)-2F_2(0)],\qquad \omega(\mathbf k)=2h(\mathbf k)/(\rho\Gamma),\\
\nabla F_2(\mathbf t)=-\tfrac12\mathbf tF_1(\mathbf t),\qquad
\nabla F_3(\mathbf t)=-\tfrac14\mathbf tF_2(\mathbf t).
\end{gathered}
\tag{S48}\label{eq:soft-transforms}
\]

Each quartic term vanishes at a constant soft leg. With \(D_R=|s-s'|^2+a_c^2\), let \(G_J,G_A,G_D\) be its soft derivative from \(JJ'/D_R\), \(A_\nabla|\Delta w|^2/D_R^2\), and \(|\Delta w|^4/D_R^3\), before multiplying by \(K_E\). Grouping their sum by transform gives the cancellation
\[
\begin{aligned}
(G_J-G_A+G_D)_{F_1}&=-\tfrac12\mathbf Q[Q^2F_1(\mathbf Q)-k^2F_1(\mathbf k)-l^2F_1(\mathbf l)],\\
(G_J-G_A+G_D)_{F_2}&=-\mathbf Q[F_2(\mathbf Q)-F_2(\mathbf k)-F_2(\mathbf l)+F_2(0)],\\
\left.\nabla_{\mathbf p}V_w\right|_0&=-\tfrac12\mathbf Q[h(\mathbf Q)-h(\mathbf k)-h(\mathbf l)].
\end{aligned}
\tag{S49}\label{eq:soft-cancellation}
\]

The finite terms can be checked explicitly. Writing \(L_{a_c}=\ln(a_c/2)+\gamma_E\) and now \(c=\mathbf k\cdot\mathbf l\), S1 gives
\[
\begin{aligned}
h(\mathbf k)&=-\pi K_E k^2[L_{a_c}+\ln k+\tfrac12]+o(1),\\
\left.\nabla_{\mathbf p}\mathcal F\right|_0
&=\tfrac12\mathbf Q[c+Q^2\ln Q-k^2\ln k-l^2\ln l].
\end{aligned}
\tag{S50}\label{eq:soft-finite-part}
\]

The dispersion constant \(1/2\), supplied by \(F_2\), fixes the nonlogarithmic \(c\) term. Soft transforms start at \(p^2\ln p\) and \(p^4\ln p\); nonzero hard transfers are smooth. Thus, at fixed cutoff,
\[
V_{{\rm can},a_c}
=-\frac{\mathbf p\cdot\mathbf Q}{\rho\Gamma}
[\omega(\mathbf Q)-\omega(\mathbf k)-\omega(\mathbf l)]
+O_{a_c,\rm hard}\!\left(p^2[1+|\ln(p/k_h)|]\right).
\tag{S51}\label{eq:soft-relation}
\]

On a smooth exactly resonant path,
\[
\omega(\mathbf Q)-\omega(\mathbf k)-\omega(\mathbf l)
=\omega(\mathbf Q)-\omega(\mathbf Q-\mathbf p)-\omega(\mathbf p)=O(p).
\]

The hard dispersion is smooth and \(\omega(\mathbf p)=O_{a_c}(p^2[1+|\ln p|])\), so the vertex has the displayed \(O(p^2[1+|\ln p|])\) order. The cutoff limit requires paired subtractions \(\widehat V=V_w-\pi K_EBL_{a_c}\), \(\widehat h=h+\pi K_Ek^2L_{a_c}\): the area term otherwise leaves \(-\pi K_EcL_{a_c}p^2\). Simultaneous soft/forward limits and microscopic GP matching in this limit are not established here.

\hypertarget{supplemental-references}{%
\section{Supplemental references}\label{supplemental-references}}

S1. P. M. Goldbart and F. Bora, \emph{Quantized vortices and superflow in arbitrary dimensions: Structure, energetics and dynamics}, \href{https://doi.org/10.1088/1751-8113/42/18/185001}{J. Phys. A \textbf{42}, 185001 (2009)}.

S2. B. N. Shashikanth, \emph{Vortex dynamics in \(\mathbb{R}^4\)}, \href{https://doi.org/10.1063/1.3673800}{J. Math. Phys. \textbf{53}, 013103 (2012)}.

S3. B. Khesin, \emph{Symplectic structures and dynamics on vortex membranes}, \href{https://doi.org/10.17323/1609-4514-2012-12-2-413-434}{Mosc. Math. J. \textbf{12}, 413--434 (2012)}.

S4. E. P. Gross, \emph{Structure of a quantized vortex in boson systems}, \href{https://doi.org/10.1007/BF02731494}{Nuovo Cimento \textbf{20}, 454--477 (1961)}.

S5. L. P. Pitaevskii, \emph{Vortex lines in an imperfect Bose gas}, \href{https://jetp.ras.ru/cgi-bin/dn/e_013_02_0451.pdf}{Sov. Phys. JETP \textbf{13}, 451--454 (1961)}.

S6. D. A. Takahashi, M. Kobayashi, and M. Nitta, \emph{Nambu-Goldstone modes propagating along topological defects: Kelvin and ripple modes from small to large systems}, \href{https://doi.org/10.1103/PhysRevB.91.184501}{Phys. Rev.~B \textbf{91}, 184501 (2015)}.

S7. M. Trippenbach, Y. B. Band, and P. S. Julienne, \emph{Theory of four-wave mixing of matter waves from a Bose-Einstein condensate}, \href{https://doi.org/10.1103/PhysRevA.62.023608}{Phys. Rev.~A \textbf{62}, 023608 (2000)}.

S8. Q. Yang, J. T. Seo, S. Creekmore, D. A. Temple, P. Ye, C. Bonner, M. Namkung, S. S. Jung, and J. H. Kim, \emph{Nonlinear phase mismatch and optimal input combination in atomic four-wave mixing in Bose-Einstein condensates}, \href{https://doi.org/10.1103/PhysRevA.67.013603}{Phys. Rev.~A \textbf{67}, 013603 (2003)}.

S9. R.-M. Hervé and M. Hervé, \emph{Étude qualitative des solutions réelles d'une équation différentielle liée à l'équation de Ginzburg-Landau}, \href{https://doi.org/10.1016/S0294-1449(16)30182-2}{Ann. Inst. H. Poincaré Anal. Non Linéaire \textbf{11}, 427--440 (1994)}.

S10. M. A. Mojahed and T. Brauner, \emph{Nonrelativistic effective field theories with enhanced symmetries and soft behavior}, \href{https://doi.org/10.1007/JHEP03(2022)086}{J. High Energy Phys. \textbf{03} (2022), 086}.

\endgroup
\end{document}